\documentclass[a4paper,12pt]{article}            
\usepackage{jheppub}
\pdfoutput=1
\usepackage[T1]{fontenc}                    
\usepackage[utf8]{inputenc}                 
\usepackage[english]{babel}                 
\usepackage{microtype}                      
\usepackage[table]{xcolor}

\usepackage{amssymb}                
\usepackage{amsmath}                
\usepackage{amsthm}                 
\usepackage{mathrsfs}               
\usepackage{mathtools}              
\usepackage{dsfont}

\newcommand*\dif{\mathop{}\!\mathrm{d}}             
\newcommand{\vect}[1]{\boldsymbol{\mathbf{#1}}}     

\let\temp\varepsilon                                
\let\varepsilon\epsilon
\let\epsilon\temp

\usepackage{hyperref}     

\usepackage{graphicx}
\usepackage{subcaption}
\usepackage[utf8]{inputenc}
\usepackage{longtable}
\usepackage{tensor}
\usepackage{lmodern}
\usepackage{ulem}

\preprint{IFT-UAM/CSIC-21-81}

\title{Hydrodynamic magneto-transport in holographic charge density wave states}
\author[1,2]{Andrea Amoretti,}
\author[3,4]{Daniel Are\'an,}
\author[1]{Daniel K. Brattan,} 
\author[1,2]{Luca Martinoia.}

\emailAdd{andrea.amoretti@ge.infn.it}
\emailAdd{daniel.arean@uam.es}
\emailAdd{danny.brattan@gmail.com}
\emailAdd{luca.martinoia@ge.infn.it}

\affiliation[1]{Dipartimento di Fisica, Universit\`a di Genova,
via Dodecaneso 33, I-16146, Genova, Italy}
\affiliation[2]{I.N.F.N. - Sezione di Genova, via Dodecaneso 33, I-16146, Genova, Italy}
\affiliation[3]{Departamento de F\'isica Te\'orica, Universidad Aut{\'o}noma de Madrid, Campus de Cantoblanco, 28049 Madrid, Spain}
\affiliation[4]{Instituto de F\'\i sica Te\'orica UAM/CSIC, Calle Nicol\'as Cabrera 13-15, 28049 Madrid, Spain}

\abstract{We employ hydrodynamics and gauge/gravity to
		study magneto-transport in phases of matter where translations are broken
	(pseudo-)spontaneously.	First we provide a hydrodynamic description of systems where translations are broken homogeneously at  nonzero lattice pressure and  magnetic field. This allows us to determine analytic expressions for all the relevant transport coefficients.	 Next we construct holographic models of those phases  and determine all the DC conductivities in terms of the dual black hole geometry. Combining the hydrodynamic and holographic descriptions we obtain analytic expression for the AC thermo-electric correlators. These are fixed in terms of the black hole geometry and a pinning frequency we determine numerically. We find an excellent
	 agreement between our hydrodynamic and holographic descriptions and show that
	 the holographic models are good avatars for the study of magneto-phonons. }

\begin{document}

\maketitle

\section{Introduction}
In recent years holographic models in which translations are broken either spontaneously or pseudo-spontaneously have been intensively studied. These models include Q-lattices \cite{Amoretti:2016bxs,Amoretti:2017frz,Amoretti:2017axe,Amoretti:2018tzw,Donos:2018kkm,Donos:2019hpp,Amoretti_2019,Amoretti:2019kuf,Amoretti:2020ica,Donos:2021ueh}, massive gravity \cite{Amoretti:2014zha,Baggioli:2014roa,Amoretti:2014mma,Amoretti:2015gna,Amoretti:2017xto,Alberte:2017oqx,Alberte:2017cch,Ammon:2019wci,Baggioli:2020edn,Baggioli:2021xuv} and phases with spatially modulated charge density \cite{Donos:2011qt,Andrade:2017cnc,Andrade:2020hpu}.

On a parallel side, hydrodynamic models have been developed which describe the physical behavior, at late times and large distances, of charged fluids in the presence of (pseudo)-Goldstone modes related to the (pseudo)-spontaneous breaking of translations \cite{Delacretaz:2016ivq,Delacretaz:2017zxd,Delacretaz:2019wzh,amoretti:hydrodynamicmagnetotransport,Armas:2019sbe,Armas:2020bmo}. The dynamics predicted by these effective theories has been tested, in the appropriate regime, against holographic models and been found to concur with a great accuracy.

Of particular relevance for experiments, is the study of the thermo-electric transport properties of these systems in the presence of an external magnetic field, as this is a very common feature in many experimental setups testing the transport behavior of strongly coupled condensed matter materials. This includes for example high-temperature superconductors \cite{Delacretaz:2016ivq,Amoretti:2019buu}. Along this line, hydrodynamic models describing the magneto-transport properties of these systems in vacua that minimize the free energy have been developed in \cite{Delacretaz:2019wzh,amoretti:hydrodynamicmagnetotransport}. On the holographic side, some of the consequences of the presence of an external magnetic field on the (pseudo)-spontaneous breaking of translations have been analyzed in the massive gravity context in \cite{Baggioli:2020edn} and in Q-lattices in \cite{Donos:2021ueh}. Still a complete study of the transport properties of these holographic models in the hydrodynamic regime with an external magnetic field is missing.

In this paper we generalize the holographic Q-lattice model with a (pseudo)-spontaneous symmetry breaking of translation, initially described in \cite{Amoretti:2017frz,Amoretti:2017axe}, in order to include an external magnetic field. We will take this magnetic field to be non-zero in the thermodynamics and thus at the same order as the chemical potential in the derivative counting of hydrodynamics. We analyze the model's thermo-electric transport properties. As already observed in \cite{Amoretti:2017frz,Amoretti:2017axe,Amoretti:2018tzw,Amoretti:2019kuf}, these kind of models correctly describe the physical behavior of systems exhibiting a spontaneous or pseudo-spontaneous symmetry breaking of translations even though they are metastable; namely, the background vacuum does not minimize the free energy. Due to this fact, it was pointed out in \cite{Armas:2019sbe,Ammon:2019apj,Armas:2020bmo,Ammon:2020xyv} that existing hydrodynamic approaches (see e.g. \cite{Delacretaz:2016ivq,Delacretaz:2017zxd} and references therein) do not apply straightforwardly to these models, and a more general formulation must be followed \cite{Armas:2019sbe}. This is due to the appearance of an additional term - the lattice pressure - in the thermodynamics. In stable systems which minimize the free energy this lattice pressure, $P_{l}$, vanishes \cite{Donos:2018kkm,Armas:2020bmo} and its thermodynamic derivatives ($\partial_{\mu} P_{l}$, $\partial_{T} P_{l}$) can be absorbed into re-definitions of the transport coefficients. In such situations we can then employ the expressions of \cite{amoretti:magnetothermaltransporta}. This will not be the case here and thus one of the main results of the paper is the generalization of the formalism of \cite{Armas:2019sbe} to include the effects of pseudo-spontaneous symmetry breaking and an external magnetic field. Moreover, by combining the hydrodynamic correlators obtained in this way with the method outlined in \cite{amoretti:magnetothermaltransporta,amoretti:hydrodynamicmagnetotransport}, we will provide an expression for the hydrodynamic AC correlators in terms of the DC conductivities of the model. The latter quantities can be computed analytically for the model at hand and have been known for a long time \cite{Amoretti:2015gna,Blake:2015ina}. Thus, we can eventually provide an analytic expression for the holographic AC thermo-electric correlators in terms of the horizon data up to one coefficient, the pinning frequency, which has to be determined numerically. Finally, we show that our analytic result is in excellent agreement with the numerically computed holographic correlators.

The paper is organized as follows. In Section \ref{sec:2} we generalize the hydrodynamic method of \cite{Armas:2019sbe} to take into account pseudo-spontaneous symmetry breaking of translations and an external magnetic field. Combining the results with the method outlined in \cite{amoretti:magnetothermaltransporta,amoretti:hydrodynamicmagnetotransport} we provide a closed form for the AC hydrodynamic correlators which depends solely on their DC values and the pinning frequency. In Section \ref{sec:holomodel} we compute the same correlators in the holographic Q-lattice model with an external magnetic field. Using the known analytic results for the DC conductivities we provide a closed form for the thermo-electric correlators in terms of horizon data and one undetermined parameter, the pinning frequency, which we determine numerically finding excellent agreement with the expected result. Finally we comment upon some features of the model and we conclude the paper in Section \ref{sec:conclusions}.

\section{Broken translation invariance with non-zero lattice pressure}
\label{sec:2}

{\ The equations of motion for the (almost-)conserved hydrodynamic charges are the conservation equations for the stress tensor and the charge current. For our system, which includes the presence of translation breaking scalar operators $O_{I}$, these take the form
    \begin{eqnarray}
        \label{Eq:ConservationEquations2}
        \partial_{\mu} \langle T^{\mu \nu}\rangle = F^{\nu \mu}\langle  J_{\mu}\rangle
         - \left(\partial^{\nu} \Phi^{I}(x_i)\right) \langle O_{I}\rangle \; , \qquad \partial_{\mu} \langle J^{\mu}\rangle
         = 0 \, , 
    \end{eqnarray}
where $T^{\mu \nu} = \langle T^{\mu \nu}\rangle$ is the stress tensor, $F^{\mu \nu}$ an external electromagnetic field strength, $J^{\mu} = \langle J^{\mu}\rangle$ a $U(1)$ charge current and $\Phi^{I}(x)$ are spatially modulated sources for the scalars $O_{I}$. In addition we will need the ``Josephson relation'' which can be thought of as generating the evolution of the translation breaking scalars. This latter relation must be derived order by order in derivatives and doing so in the presence of an external magnetic field, and in the case of explicit breaking an additional non-zero phase relaxation, is one of the main thrusts of this section.}

\subsection{Homogeneity and the Ward identities}

{\ The conservation equations \eqref{Eq:ConservationEquations2} allow one to consider a broad range of translation breaking scenarios. Here we will restrict ourselves to systems where the vev of the operators $O_{I}$ are proportional to the spatial coordinates, $\langle O_{I} \rangle \propto x^{i} \delta_{iI}$ (in the spontaneous case). On the other hand, for explicit breaking, we take the source to be proportional to the spatial coordinates $\Phi_{I}(x) = \varphi x^{i} \delta_{iI}$ with $\varphi$ a constant. Consequently the space-time derivative of the sources in the ground state of our system are constants.  This symmetry breaking pattern can be realised in models with spatial translation invariance and where the scalar operators $O_{I}$ have a constant shift symmetry such that the diagonal subgroup of this pair of symmetries remains unbroken. This ensures homogeneity of our equations of motion. Indeed, this kind of breaking has been extensively considered in the literature as a description of various systems including lattice phonons \cite{Leutwyler:1993gf,Nicolis:2017eqo}, classifications of solid state phases \cite{Nicolis:2015sra} and hydrodynamic \cite{Delacretaz:2017zxd,Delacretaz:2016ivq} and holographic \cite{Andrade:2017cnc,Alberte:2017oqx,Amoretti:2017frz,Amoretti:2017axe,Alberte:2017cch,Amoretti:2018tzw,Andrade:2018gqk,Amoretti:2019kuf,Amoretti_2019,Ammon:2020xyv,Amoretti:2020ica,Andrade:2020hpu,Baggioli:2020nay,Baggioli:2020edn,Baggioli:2021xuv} constructions of charge density wave state effective field theories.}

{\ Regarding the constant $\varphi$ setting the value of the
	scalar operator source,
	we can use it to qualitatively classify the explicitly broken regime into two cases: pseudo-spontaneous and truly explicit. The first case consists of situations where $\varphi \ll |\partial_i \langle O_i \rangle|$ so that the Goldstone bosons of spontaneously broken translation invariance have acquired a small mass and can be thought of as pseudo-Goldstone bosons. This small mass is called the ``pinning frequency'' which we shall denote by $\omega_{0}^2$. On the other hand, a truly explicit case occurs when $\varphi \gtrsim |\partial_i \langle O_i \rangle|$. This happens for example in the models of \cite{Andrade:2013gsa}, where only the source is non-zero and the vev is vanishing.}

{\ With these restrictions and definitions in place, we note that from the one-point function Ward identities \eqref{Eq:ConservationEquations2} we can also derive Ward identities for the two-point functions. In particular, imposing homogeneity of the source term for the scalar operators $\partial_i \Phi^I= \varphi \delta_i^I$, and employing the convention
    \begin{eqnarray}
     f(t,\vec{x}) = \int \frac{d^{2}k d\omega}{(2 \pi)^3} f(\omega,\vec{k}) e^{- i ( \omega t - i \vec{k} \cdot \vec{x} )} \; ,
    \end{eqnarray}
one finds at zero-wavevector ($\vec{k}=0$) the following relations
    \begin{subequations}
        \label{Eq:2ptWard}
        \begin{eqnarray}
            \label{Eq:ExplicitWardIdentity1}
            i \omega \langle Q^{i} Q^{j} \rangle &=& - \left( i \omega \mu \delta\indices{^{i}_{k}} - F\indices{^{i}_{k}} \right) \langle Q^{k} J^{j} \rangle + \varphi \langle Q^{i} O^{J} \rangle \delta^{jJ}
            - i \omega \left( \chi_{\pi \pi} - \mu n \right) \delta^{ij} \; , \qquad \\
            \label{Eq:ExplicitWardIdentity2}
            i \omega \langle Q^{i} J^{j} \rangle &=& - \left( i \omega \mu \delta\indices{^{i}_{k}} - F\indices{^{i}_{k}} \right) \langle J^{k} J^{j} \rangle + \varphi\langle J^{i} O^{J} \rangle \delta^{jJ} - i \omega n \delta^{ij} \; , \\
            \label{Eq:ExplicitWardIdentity3}
            i \omega \langle Q^{i} O^{J} \rangle &=& - \left( i \omega \mu \delta\indices{^{i}_{k}} - F\indices{^{i}_{k}} \right) \langle J^{k} O^{J} \rangle - \varphi\langle O^{I} O^{J} \rangle \delta^{iI} + \delta^{iJ} \; ,
        \end{eqnarray}
    \end{subequations}
where $Q^{i} = T^{it} - \mu J^{i}$ is the canonical heat current. These will be crucial in our second aim, deriving analytic expressions for the hydrodynamic transport coefficients, following the approach of \cite{amoretti:magnetothermaltransporta} and \cite{amoretti:hydrodynamicmagnetotransport}.}

{\ In brief, the method for generating these analytic expressions for the transport coefficients relies on the existence of  a ladder structure in \eqref{Eq:2ptWard} which reduces the number of independent correlators from six to three: $\langle O^{I} O^{J} \rangle$, $\langle J^{i} O^{J} \rangle$ and $\langle J^{i} J^{j} \rangle$. As a consequence one finds that the leading terms in the $\omega\to0$ limit
	of the original six correlators are all contained in the low frequency expansion of the independent correlators. Hence, knowing the DC values of all the correlators is as good as knowing the low frequency expansion of the independent correlators. Comparing the hydrodynamic expressions at low frequencies with what is imposed by the Ward identities \eqref{Eq:2ptWard} allows us to fix the hydrodynamic transport coefficients analytically when we know the DC terms analytically (such as in our holographic model).}

\subsection{Spontaneous case}

{\ Given that the formalism is slightly simpler we shall first consider the case of spontaneous breaking ($\varphi=0$). In the spontaneous case fluctuations of the scalar operators $O^I$ correspond to the Goldstone modes of spontaneously broken translation symmetry.}

\subsubsection{Constitutive relation}

{\ We employ the formalism of \cite{Armas:2020bmo}, making minor appropriate changes to account for the presence of a magnetic field.  The indices $I$ represent coordinates on the unbroken $ISO(2)$ manifold (the ``crystal'' in the terminology of \cite{Armas:2019sbe,Armas:2020bmo}) and we define $e^I_{\mu}=\partial_{\mu}O^I$ to be the pullback map from the $(2+1)$-dimensional spacetime onto this $2$-dimensional ``crystal manifold''. Subsequently we can define an inverse metric on the crystal $h^{IJ}=g^{\mu\nu}e^I_{\mu}e^J_{\nu}$ using the inverse spacetime metric $g^{\mu \nu}$. The tensor $h^{IJ}$ can be used to raise crystal indices. We further adopt the convention of \cite{Armas:2020bmo} and define the lower index tensor $h_{IJ}$ by $h_{IJ}=(h^{-1})_{IJ}$. Crystal indices will be lowered with respect to $h_{IJ}$.}

{\ The non-linear strain tensor $u_{IJ}$  measures the distortions of the crystal from a reference ``rest'' configuration denoted $\mathds{h}_{IJ}$. This non-linear strain tensor is defined as the difference between $h_{IJ}$ and this reference configuration i.e.~$u_{IJ}=(h_{IJ}-\mathds{h}_{IJ})/2$. For our purposes we choose the reference configuration $\mathds{h}_{IJ}=\delta_{IJ}/\alpha^2$ because it respects homogeneity and spatial isotropy. We interpret the constant parameter $\alpha$ to be the inverse size of the crystal. We make a choice to set it to one in the following since situations where $\alpha\neq1$ can be obtained by a trivial rescaling of the fields $O^I\rightarrow \alpha O^I$ \cite{Armas:2019sbe}.}

{\ Assuming small strain we can construct the free energy $F$ of the crystal plus fluid order by order in the amplitude of $u_{IJ}$. This free energy is the integral over the total pressure $F=\int\dif^2x\sqrt{-g}P$ with the total pressure up to and including quadratic terms in the strain given by
\begin{equation}\label{eqn:pressure}
	P=P_f - m B +P_l\left(u^I_I+u^{IJ}u_{IJ}\right)-\frac{1}{2}K\left(u^I_I\right)^2-G\left(u^{IJ}u_{IJ}-\frac{1}{2}\left(u^I_I\right)^2\right)+\mathcal{O}(u^3) \; . 
\end{equation}
In the above expression, $P_f$ is the thermodynamic fluid pressure, $m$ the magnetisation density, $P_l$ the lattice pressure and $K$ and $G$ are respectively the bulk and shear modulus. This should be compared to the total pressure $P$ as reported in \cite{amoretti:hydrodynamicmagnetotransport} with the major difference between our current system and those considered in \cite{amoretti:hydrodynamicmagnetotransport} being the non-zero lattice pressure ($P_{l}$) term.}

{\ With the free energy to hand \eqref{eqn:pressure} we can now order by order in derivatives construct the constitutive relations. To keep our notation compact we will use the projectors $P^{\mu\nu}=g^{\mu\nu}+u^{\mu}u^{\nu}$ and $P^{I\mu}=P^{\mu\nu}e^I_{\nu}$ and define the electric field by $E_{\mu}=F_{\mu\nu}u^{\nu}$. From here, the constitutive relations for an isotropic fluid in the Landau frame are
    \begin{subequations}
        \label{eqn:constitutive_relations}
        \begin{eqnarray}
        \label{eqn:constitutive_relations1}
        J^{\mu}&=&n u^{\mu}-P^{I\mu}\sigma_{IJ}P^{J\nu}\left(T\partial_{\nu}\frac{\mu}{T}-E_{\nu}\right)-P^{I\mu}\gamma_{IJ}u^{\nu}e^J_{\nu} \; , \\
        \label{eqn:constitutive_relations2}
        T^{\mu\nu}&=&\left(\epsilon+P\right)u^{\mu}u^{\nu}+Pg^{\mu\nu}-r_{IJ}e^{I\mu}e^{J\nu}-P^{I(\mu}P^{J\nu)}\eta_{IJKL}P^{K(\rho}P^{L\sigma)}\nabla_{\rho}u_{\sigma} \; , \qquad
        \end{eqnarray}
    \end{subequations}
where $P$ is the total pressure of \eqref{eqn:pressure}, $\epsilon$, $n$ and $s$ are the total energy, charge and entropy densities and $r_{IJ}$ is a thus-far undetermined quantity - the elastic stress tensor. These quantities are related by the thermodynamic relations 
    \begin{eqnarray}
        \dif P=s\dif T+n\dif\mu + m \dif B +\frac{1}{2}r_{IJ}h^{IJ} \; , \qquad \epsilon+P=sT+n\mu \; , 
    \end{eqnarray}
which defines $r_{IJ}$ in terms of the derivative of $P$. The total charge density and entropy can further be decomposed into free quantities, given by variation of $P_{f}$ with respect to the thermodynamic parameters $\dif P_f = s_f\dif T+n_f\dif\mu + m \dif B$, and the lattice contributions $\dif P_l=s_l\dif T+n_l\dif\mu$. We will assume that both the free and lattice thermodynamic quantities have formally similar integrated first laws, up to contribution of the magnetisation, i.e.~$\epsilon_f+P_f=s_fT+n_f\mu + mB$ and  $\epsilon_l+P_l=s_lT+n_l\mu$.}

{\ In addition to the constitutive relations for the (almost-)conserved currents $J^{\mu}$ and $T^{\mu \nu}$, as discussed above, we must supply an evolution equation for the crystal (or Goldstone) fields. At zeroth order in hydrodynamics such equations correspond to constancy of the scalars along a fluid worldline. At first order in derivatives one finds 
    \begin{equation}
        \label{eqn:configuration_equation}
        \sigma^{\phi}_{IJ}u^{\mu}e^I_{\mu}+\gamma'_{JK}P^{K\mu}\left(T\partial_{\mu}\frac{\mu}{T}-E_{\mu}\right)+\nabla_{\mu}\left(r_{JK}e^{K\mu}\right)=K^{\text{ext}}_J\;,
    \end{equation}
where $\sigma_{IJ}$, $\gamma_{IJ}$, $\gamma'_{IJ}$, $\sigma^{\phi}_{IJ}$ and $\eta_{IJKL}$ are all dissipative transport matrices and $K^\text{ext}_J$ is an external background source coupled to $O^I$ that will be zero in global thermodynamic equilibrium. We will restrict ourselves to models which have time-reversal symmetry, such that the Onsager relation $\langle O^I J^j\rangle=-\langle J^i O^J\rangle$ requires that $\gamma'_{IJ}=-\gamma_{IJ}$ \cite{Armas:2019sbe}.}

{\ The form of the constitutive relations \eqref{eqn:constitutive_relations} is the same for all values of $u_{IJ}$ - in particular one can in principle use the complete pressure and not its small amplitude expansion \eqref{eqn:pressure}. However, in practice, it is sufficient to consider fluctuations about a state of global thermodynamic equilibrium and linearise in small amplitude $u_{IJ}$. In this small strain regime the transport coefficient matrices can be taken to be strain independent at first order in derivatives. Additionally, in the presence of a constant, background, external magnetic field, we must allow for the possibility of Hall transport coefficients \cite{amoretti:magnetothermaltransporta} and subsequently decompose the transport matrices as 
\begin{equation}\label{eqn:transport_coef_decomposition}
	\left(\gamma,\sigma,\sigma^{\phi}\right)_{IJ}=\left(\gamma,\sigma,\sigma^{\phi}\right)_{(\mathrm{L})}\delta_{IJ}+\left(\gamma,\sigma,\sigma^{\phi}\right)_{(\mathrm{H})}F_{IJ}\;,
\end{equation}
where we have defined $F_{IJ}=F_{\mu\nu}e^{\mu}_Ie^{\nu}_J$. As we decompose with respect to $F_{IJ}$ rather than $\epsilon_{IJ}$ the Hall coefficients are spatial parity invariant. We identify $\sigma_{(\mathrm{L},\mathrm{H})}$ to be the longitudinal and Hall components of the charge conductivity, $\sigma^\phi_{(\mathrm{L},\mathrm{H})}$ to be the crystal diffusivity components and $\gamma_{(\mathrm{L},\mathrm{H})}$ the mixed scalar-charge conductivities. In principle we could also decompose $\eta_{IJKL}$ in terms of longitudinal and Hall bulk and shear viscosities, but because we are only interested in the diffusive sectors at zero wave-vector the viscous terms will not be relevant.}

\subsubsection{AC conductivities}

{\ We will fluctuate about a flat background $g_{\mu\nu}=\eta_{\mu\nu}$, with a constant magnetic field $F^{12}=B$ and vanishing external source for the Goldstone field $K^{\text{ext}}_I=0$. The corresponding equilibrium configuration has a fixed temperature $T=T_0$ and chemical potential $\mu=\mu_0$, no spatial velocity $u^{\mu}=(1,\vect{0})$ and a uniform value for the scalars $O^I=x^I$. We linearise the constitutive relations around this equilibrium configuration\footnote{Note that the sign of the $\delta O^{I}$ is the opposite to that used in \cite{amoretti:hydrodynamicmagnetotransport} to match the conventions of \cite{Armas:2020bmo}.},
\begin{align}\label{eqn:fluctuations}
	T&\rightarrow T_0+\delta T\;,
	&\mu&\rightarrow\mu_0+\delta\mu\;,\nonumber\\
	u^{\mu}&\rightarrow (1,v^i)\;,
	& O^I&\rightarrow x^I-\delta O^I\;,
\end{align}
and find the two-point functions by solving the non-conservation equations for fluctuations of the hydrodynamic variables in the presence of plain wave sources for the $U(1)$ field strength $\delta F^{0i}\sim\exp(-i\omega t+ik_jx^j)$ and the source terms $\delta K^{\text{ext}}_I$ and $\delta g_{\mu\nu}$. Following this procedure allows us to find the independent correlators: $\langle J^iJ^j\rangle$, $\langle J^iO^J\rangle$ and $\langle O^IO^J\rangle$. Subsequently, by applying the Ward identities \eqref{Eq:2ptWard} we can derive expressions for correlators involving the heat current\footnote{The interested reader could also readily obtain some correlators involving the thermodynamic heat current by computing the variation of the $U(1)$ charge current and scalar with respect to metric.}. This approach should be contrasted to the more standard Martin-Kadanoff method which is made difficult in the presence of lattice pressure due to the appearance of double time and mixed space/time derivatives in the constitutive relations (this is hidden in the $r_{IJ}$ containing terms of \eqref{eqn:constitutive_relations1} and \eqref{eqn:configuration_equation}).}

{\ We define the following AC conductivities in the spontaneous case\footnote{We explicitly include an argument in the AC conductivities of \eqref{Eq:DefinitionsofDC1} and \eqref{Eq:DefinitionsofDC2} to differentiate them from hydrodynamic transport coefficients as there is some notational overlap between such quantities in the literature e.g.~the hydrodynamic electric conductivity $\sigma^{ij}$ and the AC electric conductivity $\sigma^{ij}(\omega)$.}
    \begin{eqnarray}
     \label{Eq:DefinitionsofDC1}
     \left(\sigma^{ij}, \alpha^{ij}, \gamma^{iJ}\right)(\omega) &=& \left( \frac{1}{i\omega} \langle J^{i} J^{j} \rangle , \frac{1}{i\omega} \langle Q^{i} J^{j} \rangle,  \langle J^{i} O^{J} \rangle \right) \; , \\
     \label{Eq:DefinitionsofDC2}
     \left(\kappa^{ij}, X^{IJ}, \theta^{iJ}\right)(\omega) &=& \left( \frac{1}{i\omega} \langle Q^{i} Q^{j} \rangle , i \omega \langle O^{I} O^{J} \rangle,  \langle Q^{i} O^{J} \rangle \right) \; . 
    \end{eqnarray}
We have split them into two sets; the leading low frequency terms of the first set \eqref{Eq:DefinitionsofDC1} are fixed by symmetry \cite{amoretti:magnetothermaltransporta,amoretti:hydrodynamicmagnetotransport} and are the same for every system satisfying our general assumptions. The second set \eqref{Eq:DefinitionsofDC2} depend on the specific microscopic theory of the system. With that said, due to the ladder nature of the Ward identities \eqref{Eq:2ptWard}, the arbitrary frequency values of $\alpha^{ij}(\omega)$, $\kappa^{ij}(\omega)$ and $\theta^{iJ}(\omega)$ can all be derived from $\sigma^{ij}(\omega)$ and $\theta^{iJ}(\omega)$. The coefficient $X^{IJ}(\omega)$ stands on its own at the level of the spontaneous two-point Ward identities being not related to any of the others. Hence the independent conductivities are $\sigma^{ij}(\omega)$, $\gamma^{iJ}(\omega)$ and $X^{IJ}(\omega)$ and we shall give hydrodynamic expressions for these.}

{\ Since the analytic expressions for the conductivities are rather large, we employ a matrix notation similar to the one used in \cite{amoretti:hydrodynamicmagnetotransport}. We construct the following matrices of hydrodynamic transport coefficients (first line) and AC conductivities (second line)
\begin{align}
    \label{Eq:SpontaneousTransportCoeffs}
	(\hat\sigma,\hat\sigma_{\phi},\hat\gamma)&=(\sigma,\sigma^{\phi},\gamma)_{(\mathrm{L})}\mathds{1}_2-(\sigma,\sigma^{\phi},\gamma)_{(\mathrm{H})}F\;,\\
	\label{Eq:DCtransportcoeffs}
	(\hat\sigma,\hat\alpha,\hat\kappa,\hat\gamma,\hat X,\hat\theta)(\omega)&=(\sigma,\alpha,\kappa,\gamma,X,\theta)_{(\mathrm{L})}(\omega)\mathds{1}_2-(\sigma,\alpha,\kappa,\gamma,X,\theta)_{(\mathrm{H})}(\omega)F^{-1} \; . 
\end{align}
We also define the following additional terms for notational convenience
\begin{align}
	\hat\sigma'&=\hat\gamma^2+\hat\sigma\cdot\hat\sigma_\phi&\hat\rho&=2\hat\gamma+F\cdot\hat\sigma-n_f\mathds{1}_2 \; . 
\end{align}
The coefficients of \eqref{Eq:SpontaneousTransportCoeffs} are the transport coefficients appearing in the constitutive relations \eqref{eqn:constitutive_relations}. In terms of them the three independent AC conductivities are
\begin{subequations}
    \label{Eq:ACSpontaneousconductivities}
	\begin{align}
		\hat\sigma(\omega)&=\hat\Lambda^{-1}\cdot\left[\omega P_l\left(in_f\hat\rho-\omega w_f\hat\sigma\right)+n_f^2\hat\sigma_\phi-\left(n_fF+i\omega\chi_{\pi\pi}\mathds{1}_2\right)\hat\sigma'\right]\;,\\
		\hat\gamma(\omega)&=\hat\Lambda^{-1}\cdot\left(i\omega w_f\hat\gamma-n_f\hat\sigma_\phi+F\cdot\hat\sigma'\right)\;,\\
		\hat X(\omega)&=\hat\Lambda^{-1}\cdot\left(F\cdot\hat\rho+i\omega w_f\mathds{1}_2-\hat\sigma_\phi\right)\;,\\
		\hat\Lambda&=\omega P_l\left(iF\cdot\hat\rho-\omega w_f\mathds{1}_2\right)+\hat\sigma_\phi(Fn_f-i\omega\chi_{\pi\pi}\mathds{1}_2)
		-F^2\cdot\sigma'\;,
	\end{align}
\end{subequations}
where we defined $w_f$ to be the fluid enthalpy density $w_f=\epsilon_f+P_f$. The above expressions are a result of hydrodynamics alone following from using our constitutive relations \eqref{eqn:constitutive_relations}.

{\ Undetermined in the above are the hydrodynamic transport coefficients: $\sigma_{IJ}$, $\sigma^{\phi}_{IJ}$ and $\gamma_{IJ}$. For the spontaneous case, we will write these coefficients as functions of $X_{iJ}(0)$, $\theta_{IJ}(0)$ and $\kappa_{ij}(0)$ by employing the Ward identities (as in \cite{amoretti:magnetothermaltransporta,amoretti:hydrodynamicmagnetotransport}). We find
\begin{subequations}
    \label{Eq:ACSpontaneoustransportcoeffs}
	\begin{align}
		\hat\sigma&=\hat\Phi^{-1}\cdot\left(\hat\kappa(0)+2\chi_{\pi\pi}\hat\theta(0)-\chi_{\pi\pi}^2\hat X(0)-\mu^2n_fF^{-1}\right)+n_f F^{-1} \; , \\
		\hat\sigma_{\phi}&=\hat\Phi^{-1}\cdot\left[F\cdot(P_l^2\hat X(0)-\hat\kappa(0)-2P_l\hat\theta(0))+\mu(\mu n_f-2w_f)\mathds{1}_2\right]\cdot F \; , \\
		\begin{split}
			\hat\gamma&=\hat\Phi^{-1}\cdot\left[F\cdot\left(P_l\chi_{\pi\pi}\hat X(0)+(w_f-2\chi_{\pi\pi})\hat\theta(0)-\hat\kappa(0)\right)\right.\\
			&\qquad\left.+\mu(\mu n_f-w_f)\mathds{1}_2\right] \; , 
		\end{split}\\
		\hat\Phi&=\left(\mu\mathds{1}_2-F\cdot\hat\theta(0)\right)^2+\left(F\cdot\hat\kappa(0)-\mu(\mu n_f-2\chi_{\pi\pi})\mathds{1}_2\right)\cdot F\cdot\hat X(0) \; . 
	\end{align}
\end{subequations}
The expressions \eqref{Eq:ACSpontaneousconductivities} and \eqref{Eq:ACSpontaneoustransportcoeffs} correctly reduce to the expressions obtained in \cite{amoretti:hydrodynamicmagnetotransport} at $P_l=0$ once appropriate identifications are made between the transport coefficients to account for the differing ways that the constitutive relations were constructed.}

\subsection{Pseudo-spontaneous and explicit cases}

{\ We now consider the cases of pseudo-spontaneous and explicit breaking of translation invariance by the scalars. The distinction between these two cases is somewhat loose, but we remind the reader that we identify the former as satisfying $\varphi \ll |\partial_i \langle O_i \rangle|$ while the latter consists of all other situations. Again, we will employ the formalism of \cite{Armas:2020bmo}, but now we must account for a non-zero phase relaxation in addition to the existence of an external magnetic field.}

\subsubsection{Constitutive relation}

{\ In the explicit case the procedure for constructing the effective hydrodynamic theory is broadly the same. In particular, the constitutive relations \eqref{eqn:constitutive_relations} and charge conservation equations remain unchanged modulo the inclusion of an explicit mass term for the scalar fields in the evolution equation for the spatial momentum
\begin{equation}
    \label{Eq:MomentumConservation}
	\partial_t P^i+\partial_j T^{ij}=F^{i\mu}J_{\mu}-K_I^{\text{ext}}e^{Ii}+\omega_0^2\chi_{\pi\pi}O^I\delta^{Ii} \; , 
\end{equation}
where $\omega_0^2$ is the pinning frequency. This addition follows directly from the one-point Ward identities \eqref{Eq:ConservationEquations2} when the scalars are sourced by a homogeneous and isotropic term of the form $\Phi^{I} = \varphi x^{I}$. In particular, we identify $\varphi = \omega_{0}^2 \chi_{\pi \pi}$ as in \cite{amoretti:hydrodynamicmagnetotransport}.} 

{\ While the conservation equations are mostly unchanged, to describe the evolution of the scalars accurately we must also account for the fact that they are no longer massless and will have a tendency to spread out in space-time. We can phenomenologically track this effect by adding a non-zero phase relaxation term $\Omega^{IJ} O_J$ to the Josephson relation \cite{amoretti:hydrodynamicmagnetotransport} i.e.~
    \begin{eqnarray}
        \label{eqn:configuration_equation_explicit}
        \sigma^{\phi}_{IJ}u^{\mu}e^I_{\mu}+\gamma'_{JK}P^{K\mu}\left(T\partial_{\mu}\frac{\mu}{T}-E_{\mu}\right)+\nabla_{\mu}\left(r_{JK}e^{K\mu}\right)=\Omega^{IJ} O_J + K^{\text{ext}}_J \; . 
    \end{eqnarray}
The phase-relaxation tensor $\Omega^{IJ}$, like the other hydrodynamic transport coefficients discussed here, decomposes with respect to $SO(2)$ rotation invariance and microscopic parity invariance
    \begin{equation}
        \Omega^{IJ}=\Omega_{(\mathrm{L})}\delta^{IJ}+\Omega_{(\mathrm{H})}F^{IJ} \; . 
    \end{equation}
While in principle the new phase relaxation term in \eqref{eqn:configuration_equation_explicit} is independent of the other transport coefficients, it turns out that Onsager relations impose a constraint on its value, in particular we found that
    \begin{equation}\label{eqn:onsager}
        \Omega^{IJ} = \omega_0^2\chi_{\pi\pi}\delta^{IJ}\;.
    \end{equation}
This may seem surprising given that in previous works \cite{amoretti:hydrodynamicmagnetotransport} the phase relaxation is an independent transport coefficient with a non-zero Hall term. What is missing here is that compared to the formalism of \cite{amoretti:hydrodynamicmagnetotransport}, the time evolution of the scalar field (i.e.~the first term of \eqref{eqn:configuration_equation_explicit}) is not normalised to the identity matrix. Consequently, in the present formalism, the inverse crystal diffusivity
$(\sigma^{\phi})^{-1}$
plays the same role as the phase-relaxation tensor of \cite{amoretti:hydrodynamicmagnetotransport}. To compare the results with \cite{amoretti:hydrodynamicmagnetotransport}, in the limit of vanishing lattice pressure, one must rescale \eqref{eqn:configuration_equation_explicit} with the inverse crystal diffusivity and identify the phase relaxation tensor of \cite{amoretti:hydrodynamicmagnetotransport} to be $\omega_{0}^2 \chi_{\pi \pi} (\sigma^{\phi})^{-1}$.}

\subsubsection{AC conductivities}

{\ To compute the AC conductivities and identify the hydrodynamic transport coefficients we proceed as for the spontaneous case. The global thermodynamic equilibrium configuration is unchanged and we can fluctuate again with \eqref{eqn:fluctuations} to find the linearized expressions. However, to express the results in a compact form we once more require some additional notation. To begin with we define two new AC transport terms:
    \begin{eqnarray}
     (\varpi^{iJ}, \zeta^{IJ})(\omega) = \frac{1}{i \omega} \left(\langle J^{i} O^{J} \rangle, \langle O^{I} O^{J} \rangle - \frac{1}{\varphi} \delta^{IJ}  \right) \; .
    \end{eqnarray}
Contrasted to \eqref{Eq:DefinitionsofDC1} and \eqref{Eq:DefinitionsofDC2} these contain differing overall powers of the frequency reflecting the differing behaviour of the explicit correlators at low frequencies. Subsequently we decompose our AC conductivities as new matrices
    \begin{eqnarray}
        \label{Eq:Thermalconductivities}
        (\hat\sigma,\hat\alpha,\hat\kappa,\hat\varpi,\hat\zeta)(\omega)=(\sigma,\alpha,\kappa,\varpi,\zeta)_{(\mathrm{L})}(\omega)\mathds{1}_2+(\sigma,\alpha,\kappa,\varpi,\zeta)_{(\mathrm{H})}(\omega)F \; . 
    \end{eqnarray}
The use of $F$ in the explicit case, rather than $F^{-1}$ as we used in the spontaneous case \eqref{Eq:DCtransportcoeffs}, reflects the smoothness of these latter conductivities as $B \rightarrow 0$. We also introduce one additional new quantity,
    \begin{align}\label{Gammafreq}
        \Gamma&=\omega_0^2\chi_{\pi\pi}-\omega^2P_l \; ,
    \end{align}
not to be confused with any explicit momentum loss tensor.} 

{\ With these definitions to hand we find that the three independent AC correlators are then given by
\begin{subequations}
    \label{Eq:ExplicitACconductivities}
	\begin{align}
		\hat\sigma(\omega)&=\hat\Xi^{-1}\cdot\left[\Gamma\omega w_f\hat\sigma+\omega n_f^2\hat\sigma_\phi-i(\omega^2-\omega_0^2)\chi_{\pi\pi}\hat\sigma'-n_f\left(i\Gamma\hat\rho+\omega F\cdot\hat\sigma'\right)\right]\;,\\
		\hat\varpi(\omega)&=\hat\Xi^{-1}\cdot\left(\omega w_f\hat\gamma+i(n_f\hat\sigma_\phi-F\cdot\hat\sigma')\right)\;,\\
		\hat\zeta(\omega)&=\frac{1}{\omega_0^2\chi_{\pi\pi}}\hat\Xi^{-1}\cdot\left(\omega\chi_{\pi\pi}\hat\sigma_\phi-\omega P_l\left(F\cdot\hat\rho+i\omega w_f\mathds{1}_2\right)+iF\cdot(n_f\hat\sigma_\phi-F\cdot\hat\sigma')\right)\;,\\
		\hat\Xi&=\Gamma\left(\omega w_f\mathds{1}_2-iF\cdot\hat\rho\right)+\omega n_fF\cdot\hat\sigma_\phi-i(\omega^2-\omega_0^2)\chi_{\pi\pi}\hat\sigma_\phi-\omega F^2\cdot\hat\sigma'\;. \label{denominatore}
	\end{align}
\end{subequations}
 Consequently, by employing the Ward identities \eqref{Eq:2ptWard}  the three hydrodynamic transport matrices $\hat\sigma$, $\hat\sigma_{\phi}$ and $\hat\gamma$ can be expressed in terms of the DC values of the electric, thermoelectric and thermal conductivity $\hat\sigma(0)$, $\hat\alpha(0)$ and $\hat\kappa(0)$ hydrodynamic transport coefficients: 
\begin{subequations}
	\label{Eq:ExplicitACtransportcoeffs}
	\begin{align}
		\hat\sigma&=-\hat\Psi^{-1}\cdot\hat\pi(0)\;,\\
		\hat\sigma_\phi&=\hat\Psi^{-1}\cdot\left[w_f^2\mathds{1}_2
		+\left(F\cdot\hat\pi(0)-2w_f(\hat\alpha(0)+\mu\hat\sigma(0))\right)\cdot F\right]+n_fF\;,\\
		\hat\gamma&=\hat\Psi^{-1}\cdot\left[F\cdot\hat\pi(0)-w_f(\hat\alpha(0)+\mu\hat\sigma(0))\right]+n_f\mathds{1}_2\;,\\
		\hat\Psi&=\mu^2\hat\sigma(0)+2\mu\hat\alpha(0)+\hat\kappa(0)\;,
	\end{align}
\end{subequations}
where we have defined
\begin{equation}
	\hat\pi(0)=\hat\alpha^2(0)-\hat\kappa(0)\cdot\hat\sigma(0)\;,
\end{equation}
i.e.~minus the determinant of the DC thermoelectric conductivity matrix.
Notice that, compared to the spontaneous case, in the explicit case the DC electric, thermo-electric and charge-scalar conductivities are not fixed by symmetries. This implies that the value of the transport coefficients \eqref{Eq:ExplicitACtransportcoeffs} can be in principle obtained in a real experiment by measuring only the DC electric thermo-electric and thermal conductivities.

Finally, we can again take the limit of $P_{l} \rightarrow 0$ and match these expressions, \eqref{Eq:ExplicitACconductivities} and \eqref{Eq:ExplicitACtransportcoeffs}, against those in \cite{amoretti:hydrodynamicmagnetotransport} with the appropriate identifications. The agreement is perfect.}

\section{Holographic model}
\label{sec:holomodel}


{\ To test our hydrodynamic theory in a precise scenario  we now study the transport properties of a holographic model consisting of a bidimensional `Q-lattice'~\cite{Donos:2014uba} with action
    \begin{eqnarray}
        \label{Eq:HolographicAction}
        S &=& \int d^{3+1}x \sqrt{-g}\biggl( R-V[\phi]-\frac{1}{2}(\partial\phi)^2-\frac{Z[\phi]}{4}F^2-\frac{1}{2}Y[\phi]\sum_{i=1,2}(\partial\psi_i)^2\biggr) \,. 
        \label{eq:holoact}
    \end{eqnarray}
This model is suitable for describing the symmetry breaking pattern we considered in previous sections since it enjoys a shift symmetry $\psi_i\to \psi_i+c_i$. Indeed, we will impose that
    \begin{equation}
        \label{Eq:BackgroundPsi}
        \psi_i= k x^i\,,\qquad x^i=\{x,y\}\,,
    \end{equation}
which breaks spatial translations and the shift symmetry to a diagonal $U(1)$. This form for the background values of $\psi_{i}$ allow us to find solutions where the metric, $U(1)$ gauge field and scalar $\phi$ in \eqref{eq:holoact} depend only on the radial coordinate; translations are broken homogeneously \cite{Amoretti:2017axe,Amoretti:2017frz,Amoretti:2018tzw,Amoretti:2019buu,Amoretti:2019kuf,Amoretti:2020ica}. From this point onward, given our choice of background axion in \eqref{Eq:BackgroundPsi} we will now identify capital Latin indices $I,J,K,L,\ldots$ which labeled crystal directions with the equivalent spatial indices $i,j,k,l,\ldots$.}

{\ We take as an ansatz for our background solution to the equations of motion coming from \eqref{Eq:HolographicAction} the following:
    \begin{equation}
        \label{Eq:BackgroundAnsatz1}
        ds^2 = \frac{1}{r^2}\left(-f(r)dt^2+ \frac{dr^2}{f(r)} + g(r)d\vec{x}^2 \right), \quad
        A=a(r) dt-B\,y dx\,,\quad\phi=\phi(r)\,.
    \end{equation}
The expressions in \eqref{Eq:BackgroundAnsatz1} can accommodate configurations which are asymptotic to AdS, provided we impose particular $\phi\to0$ asymptotics for the scalar couplings 
    \begin{eqnarray}\label{Eq:Asymptoticpotential}
        V[\phi] = - 6 - \phi^2 + \mathcal{O}(\phi^3) \; , \qquad Z[\phi] = 1 + \mathcal{O}(\phi) \; , \qquad Y[\phi] = \frac{\Upsilon}{2} \phi^2 + \mathcal{O}(\phi^3) \,. \qquad
    \end{eqnarray}
As has been thoroughly discussed in \cite{Amoretti:2017frz,Amoretti:2018tzw,Amoretti:2019kuf}, the asymptotic behavior of $\phi$ towards the boundary of AdS will determine whether translations are broken explicitly or spontaneously. Given the asymptotics of \eqref{Eq:Asymptoticpotential}, the scalar $\phi(r)$ behaves as 
    \begin{equation}
        \label{Eq:BackgroundAnsatz2}
        \phi(r)=\lambda r+\phi_v r^2+O(r^3) \; .
    \end{equation}
Briefly, these asymptotic choices for the scalar and axions allow one to repackage the fields ($\phi, \psi_i$) into a pair of complex scalars $\Phi_i\sim\phi\exp(i\psi_i)$ as in \cite{Donos:2013eha}. Solutions with $\lambda=0$ correspond to a theory where the operators dual to $\psi_i$ break translations spontaneously, while backgrounds with $\lambda\neq0$ break them explicitly. At zero magnetic field, both spontaneous and explicit solutions with potentials satisfying \eqref{Eq:Asymptoticpotential} have been constructed in \cite{Amoretti:2017axe,Amoretti:2017frz,Amoretti:2018tzw,Amoretti:2019buu}. In this section
we will explore such solutions further, focusing on their transport properties in the presence of an external magnetic field, and for a choice of potentials where
    \begin{eqnarray}\label{potential}
        \label{eq:holopots}
        V[\phi] = - 6 \cosh\left( \frac{\phi}{\sqrt{3}} \right) \, , \qquad Z[\phi] = \exp\left( - \frac{\phi}{\sqrt{3}} \right) \,, \qquad Y[\phi] = \left( 1 - e^{\phi} \right)^2 \; .
    \end{eqnarray}
Consequently we set $\Upsilon=2$ in \eqref{Eq:Asymptoticpotential}.}

\subsection{Summary of the thermodynamics}

{\ To match the holographic model to our hydrodynamic theory, we first need to determine the thermodynamic quantities in the boundary field theory corresponding to our backgrounds \eqref{Eq:BackgroundAnsatz1}. On the boundary, the thermodynamic state of our theory is determined by the temperature $T$, the chemical potential $\mu$, the external magnetic field $B$ and in principle the wavelength $k$ of the crystal. In our models however we will treat $k$ as an external parameter and not minimize the free energy with respect to this quantity. That this approach gives, at the level of the transport properties, the same result as a more sophisticated model (see \textit{e.g.}~\cite{Andrade:2017cnc,Andrade:2020hpu}) where $k$ is fixed by minimizing the free energy, is discussed for example in \cite{Amoretti:2017frz,Amoretti:2020ica}. 
Additionally, to specify the groundstate at the boundary we require either the vev of the scalar in the spontaneous case ($\phi_{v}$) or the boundary source ($\lambda$) in the explicit case.}

{\ As detailed in Appendix \ref{app:numerics} we construct numerical solutions corresponding to non-zero temperature and charge density states in an external magnetic field. In addition to \eqref{Eq:BackgroundAnsatz2}, one can show that the boundary behavior of the functions in our background ansatz \eqref{Eq:BackgroundAnsatz1} is of the form
	  \begin{subequations}
		\label{Eq:UVexpansionbackgroundexp}
		\begin{eqnarray}
		\label{Eq:UVexpansionbackgroundexpf}
		f(r) &=& 1 - \frac{\lambda^2}{4} r^2 - \frac{\epsilon_{f}}{6} r^3+O(r^4) \,, \\
		g(r)&=& 1 - \frac{\lambda^2}{4} r^2 - \frac{\lambda \phi_{v}}{3} r^3 +O(r^4) \,,  \\
		\label{Eq:UVexpansionbackgroundexpgauge}
		a(r) &=& \mu- n_{f} r+O(r^3) \,,
		\end{eqnarray}
	\end{subequations}
where in the gauge field expansion \eqref{Eq:UVexpansionbackgroundexpgauge} we recognise the chemical potential $\mu$, and the free electric charge density $n_{f}$. Similarly in \eqref{Eq:UVexpansionbackgroundexpf} we find the energy density $\epsilon_{f}$ appearing at some subleading order. Solutions dual to states at finite temperature present a horizon at a finite $r_h$ in the bulk. The asymptotic behavior of our background fields towards the horizon read
    \begin{eqnarray}
        \label{Eq:Nearhorizonfields}
        ds^2 &=& -4 \pi T (r_{\mathrm{h}}-r) dt^2 + \frac{dr^2}{4 \pi T (r_{\mathrm{h}}-r)} + \frac{s_{f}}{4 \pi}(dx^2+dy^2) \;, \qquad \\
        a(r) &=& a_{h,1} (r_{\mathrm{h}}-r)+..., 
        \qquad \phi = \phi_{\mathrm{h}}+... \,,
    \end{eqnarray}
where $T$ and $s_{f}$ correspond respectively to the temperature and the free entropy density of the dual system.

{\ We can make further progress in determining the thermodynamics of the system in terms of the near-horizon asymptotics of our solutions by making use of two radially-conserved quantities that follow from the background equations of motion. First, the Maxwell equation implies the conservation of
    \begin{equation}
        \label{Eq:Maxwellradial}
        - g(r) Z[\phi(r)] a'(r)\,,
    \end{equation}
which asymptotes to the free electric charge density $n_{f}$ at the boundary (hence the overall sign choice). Consequently, the leading term in the near horizon expansion of the gauge field ($a_{h,1}$) can be given in terms of the boundary electric charge density $n$,
    \begin{eqnarray}
        a_{h,1} = - \frac{4 \pi n_{f}}{s_{f} Z_{\mathrm{h}}} \; , \qquad Z[\phi_{\mathrm{h}}] = Z_{\mathrm{h}} \; .
    \end{eqnarray}
The second radially conserved quantity follows from the Einstein equations which imply
    \begin{equation}
        \label{Eq:Einsteinradial}
        \left[n_{f} a(r)+ \frac{g^2(r)}{r^2} \left(\frac{f(r)}{g(r)}\right)' - k^2 I_{Y}(r) - B^2 I_{Z}(r) \right]' = 0 \, , 
    \end{equation}
where we have introduced the integrals 
    \begin{equation}
        \label{Eq:IYIZintegrals}
        I_{Y}(r) = \int_{w=0}^{r} d w \; \frac{Y[\phi(w)]}{w^2} \,,\qquad I_{Z}(r) = \int_{w=0}^{r} d w \; \frac{Z[\phi(w)]}{g(w)} \,.
    \end{equation}
The former of these represents the thermodynamic response of the theory to broken translation invariance which is captured in the ``lattice pressure''~\cite{Armas:2019sbe} $P_{l} = - k^2 I_{Y}(r_{\mathrm{h}})$, while the latter is related to the magnetisation density, $m = - B I_{Z}(r_{\mathrm{h}})$, as discussed in~\cite{Blake:2015ina,Lucas:2015pxa}. Importantly, at the horizon and boundary \eqref{Eq:Einsteinradial} takes the following values
    \begin{eqnarray}
        \mathrm{boundary} : \mu n_{f} + \lambda \phi_{v} - \frac{\epsilon}{2} \,, \qquad \mathrm{horizon} : - s_{f} T + m B + P_{l} \; . 
    \end{eqnarray}
Equating these expressions shows that these thermodynamic quantities satisfy a Smarr-type relation
    \begin{eqnarray}
        \epsilon_{f} = 2 \left( s_{f} T  + \mu n_{f} - m B - P_{l} + \lambda \phi_{v} \right) \, . 
    \end{eqnarray}
}
 
\subsection{Conserved bulk radial currents at the fluctuation level}
\label{ssec:holoDCxp}
{\ It will be possible to compute analytically the zero frequency value of several correlators in the holographic model.  These will include our input data for the hydrodynamic model (\textit{e.g.}~the DC conductivities\footnote{See also \cite{Donos:2018kkm,Donos:2019tmo,Donos:2019hpp,Gouteraux:2018wfe} for the derivation of holographic DC quantities in analogous models.}). To do this we shall turn on a constant background electric field at the level of fluctuations and compute the response of the theory. We therefore consider the following perturbations for the gauge field and metric
    \begin{subequations}
    \label{Eq:DiffusionAnsatz}
    \begin{eqnarray}
     \delta A_{x}(r) = a_{x}(r)-p_{x}(r) t \; , \quad \delta g_{tx}(r) = \frac{1}{r^2} \left( h_{x}(r)-\tilde{p}_{x} (r) t \right) \; , \quad \delta g_{rx}(r) = \frac{1}{r} \tilde{h}_{x}(r) \; , \qquad\;\;\\ 
     \delta A_{y}(r) = a_{y}(r)-p_{y}(r) t \; , \quad \delta g_{ty}(r) = \frac{1}{r^2} \left( h_{y}(r)-\tilde{p}_{y} (r) t \right) \; , \quad \delta g_{ry}(r) = \frac{1}{r} \tilde{h}_{y}(r) \; , \qquad \;\;\;
    \end{eqnarray}
    \end{subequations}
where the radial functions $p_{x}(r)$, $p_{y}(r)$, $\tilde{p}_{x}(r)$ and $\tilde{p}_{y}(r)$ will correspond to turning on a small electric field at the fluctuation level. The difference between the spontaneous and explicit case will be encoded in the axion fields. In the former case they are taken to have the form
    \begin{eqnarray}
     \delta \psi_{x}(r) = \frac{\chi_{x}(r)}{r} - k \delta V_{x} t  \; , \qquad \delta \psi_{y}(r) = \frac{\chi_{y}(r)}{r} - k \delta V_{y} t \; , 
    \end{eqnarray}
where $\delta V_{x}$ and $\delta V_{y}$ are sliding modes which encode an ambiguity in the definition of the vev of the axions at the boundary \cite{Davison:2015taa,Donos:2018kkm,Amoretti:2017frz,Donos:2019hpp}. This ambiguity is fixed by conditions at the horizon. Meanwhile in the explicit case the axions are purely radial functions
    \begin{eqnarray}
     \delta \psi_{x}(r) = \chi_{x}(r)  \; , \qquad \delta \psi_{y}(r) = \chi_{y}(r) \; ,
    \end{eqnarray}
with no sliding mode ambiguity.}

{\ The explicit time dependence of the ans\"{a}tze \eqref{Eq:DiffusionAnsatz} will drop out from the linearised equations of motion provided that $p_{x}(r)$, $p_{y}(r)$, $\tilde{p}_{x}(r)$ and $\tilde{p}_{y}(r)$ take particular forms:
    \begin{subequations}
    \label{Eq:p1p2rsols}
    \begin{eqnarray}
        p_{x}(r)= p_{x}^{(0)} + n_{f} \bar{E}_{x} a(r) \ , \qquad \tilde{p}_{x}(r)= -n_{f} \bar{E}_{x} f(r) \ , \\
        p_{y}(r)= p_{y}^{(0)} + n_{f} \bar{E}_{y} a(r) \ , \qquad \tilde{p}_{y}(r)= -n_{f} \bar{E}_{y} f(r) \ , 
    \end{eqnarray}
    \end{subequations}
where $p_{x}^{(0)}$, $p_{y}^{(0)}$, $\bar{E}_{x}$ and $\bar{E}_{y}$ are free constants. This perturbation represents a stationary state where the applied electric field is balanced against momentum loss at the fluctuation level. From this starting point one can construct Frobenius expansions for the fluctuation fields in the near horizon region and at the boundary. At the horizon we impose regularity conditions.}

\subsubsection{Spontaneous case}

{\ Given the perturbations we have switched on, it is relatively straightforward to massage the bulk equations of motion into sets of conservation equations for radial currents. For example, from the Maxwell equations for the gauge field perturbation one can identify the following radially conserved currents:
    \begin{subequations}
    \label{Eq:BulkChargeCurrents}
    \begin{eqnarray}
        \delta \mathcal{J}_{x}(r) &=& Z[\phi(r)] f(r) \left( a_{x}'(r) + \frac{r B}{g(r)} \tilde{h}_{y}(r) \right) - \frac{n_{f}}{g(r)} h_{x}(r) - n_{f} B \bar{E}_{y} I_{Z}(r) \; , \qquad \\
        \delta \mathcal{J}_{y}(r) &=& Z[\phi(r)] f(r) \left( a_{y}'(r) - \frac{r B}{g(r)} \tilde{h}_{x}(r) \right) - \frac{n_{f}}{g(r)} h_{y}(r) + n_{f} B \bar{E}_{x} I_{Z}(r)  \; . \qquad
    \end{eqnarray}
    \end{subequations}
At the boundary these currents encode electric charge conservation in the zero frequency limit and their form as expressed above does not depend on whether we are considering the spontaneous or explicit case. We can arrange for these bulk currents to tend to the vev of the electric charge current as $r \rightarrow 0$ i.e.~$\delta \mathcal{J}_{i}(0) =\lim_{r \rightarrow 0} \partial_r a_i(r) = \langle J^{i} \rangle$. We do not make any explicit magnetisation subtractions. This requires that we make the following identifications
    \begin{subequations}
    \label{Eq:p1p2identificationspontaneous}
    \begin{eqnarray}
        p_{x}^{(0)} &=& \left( s_{f} T - m B - k^2 I_{Y} \right) \bar{E}_{x} - \frac{B}{n_{f}} \langle J^{y} \rangle + \frac{\delta s_{x}}{n_{f}} \; , \\
        p_{y}^{(0)} &=& \left( s_{f} T - m B - k^2 I_{Y} \right) \bar{E}_{y} + \frac{B}{n_{f}} \langle J^{x} \rangle + \frac{\delta s_{y}}{n_{f}} \; , \\
        \label{Eq:p1p2identificationspontaneousvevs}
        \delta s_{i} &:=& k \phi_{v}^2 \chi_{i}(0) \; ,
    \end{eqnarray}
    \end{subequations}
in the spontaneous case where $\langle J^{i} \rangle$ is the vev of the total spatial electric charge current at the boundary and $\delta s_{i}$ is the boundary source for the Goldstone field. As the values of $p_{x}^{(0)}$ and $p_{y}^{(0)}$ were free in \eqref{Eq:p1p2rsols} this presents no difficulty. That the identification of the phonon source is correct and unambiguous was addressed in \cite{Amoretti:2018tzw,Amoretti_2019}.}

{\ There are also a pair of conserved currents related to heat transport. These have the form
    \begin{subequations}
    \label{Eq:SpontaneousBulkHeat}
    \begin{eqnarray}
        \delta \mathcal{Q}_{x}(r) &=& - f(r) \left( \frac{h_{x}(r)}{r^2} \right)' + \left( \frac{f(r)}{r^2} \right)' h_{x}(r) + \left( a(r) - \frac{B^2}{n_{f}} I_{Z}(r) \right) \langle J^{x} \rangle \qquad \nonumber \\
        &\;& + B \bar{E}_{y} \left( M_{Q}(r) - (s_{f} T + n_{f} a(r) - m B - k^2 I_{Y} ) I_{Z}(r) \right) \nonumber \\
        &\;& - \frac{B}{n_{f}} \delta s_{y} I_{Z}(r) - k^2 \delta V_{x} I_{Y}(r) \; , \qquad \\
        \delta \mathcal{Q}_{y}(r) &=& - f(r) \left( \frac{h_{y}(r)}{r^2} \right)' + \left( \frac{f(r)}{r^2} \right)' h_{y}(r) + \left( a(r) - \frac{B^2}{n_{f}} I_{Z}(r) \right) \langle J^{y} \rangle \qquad \nonumber \\
        &\;& - B \bar{E}_{x} \left( M_{Q}(r) - (s_{f} T + n_{f} a(r) - m B - k^2 I_{Y} ) I_{Z}(r) \right) \qquad \nonumber \\
        &\;& + \frac{B}{n_{f}} \delta s_{x} I_{Z}(r) - k^2 \delta V_{y} I_{Y}(r) \; , \qquad 
    \end{eqnarray}
    \end{subequations}
where we have taken\footnote{There is an ambiguity in the definition of $M_{Q}(r)$ as it appears in the bulk heat current. In particular, redefining $M_{Q}(r)$ to be the linear combination
    \begin{eqnarray}
       M_{Q}(r) &=& - 2 n_{f} B \int_{0}^{r} \frac{dw}{g(w)} \; \left( (\alpha - 1) a(w) Z[\phi(w)] - (\alpha + 1) \frac{I_{Z}(w)}{Z[\phi(w)]} \right) \; , 
    \end{eqnarray}
where $\alpha$ is an arbitrary constant, also leads to a conserved heat current that tends to the correct form at the boundary. However, it differs at the horizon and leads to a shift of the thermal Hall conductivity. We have chosen $M_{Q}$ such that $M_{Q}(r_\mathrm{h}) = M_{E} - \mu m$ where $M_{E}$ is the thermodynamic magnetisation energy.}
    \begin{eqnarray}
        M_{Q}(r) = - 2 n_{f} B \int_{0}^{r} dw \; \frac{Z[\phi(w)] a(w)}{g(w)} \; , \qquad
        M_{Q} = M_{Q}(r_{\mathrm{h}}) \; . \qquad
    \end{eqnarray}
It can be shown using the asymptotic expansions that these expressions \eqref{Eq:SpontaneousBulkHeat} tend to the canonical heat current at the boundary; again we have made no magnetisation subtractions.}

{\ Additionally there are radially conserved currents corresponding to fluctuations of the axions. These turn out to be linear combinations of the bulk heat and electric charge currents so we relegate their expressions to appendix \ref{appendix:scalarcurrent}.}

{\ Comparing the bulk electric charge current and heat currents at the horizon and boundary allows us to compute the DC conductivities in the standard manner. Firstly, it is possible to identify the boundary electric field ($E_{i}$) and thermal gradients ($\partial_{i} T/T$) from the Frobenius expansions
    \begin{eqnarray}
     \label{Eq:boundarysourceidentifications}
     E_{i} = -\lim_{r\to0} \partial_{t} \left( \delta a_{i} + \frac{\mu r^2}{f(r)} \delta g_{ti} \right) \; , \qquad \frac{\partial_{i} T}{T} = \lim_{r\to0}\partial_{t} \left( \frac{r^2}{f(r)} \delta g_{ti} \right) \; . \qquad 
    \end{eqnarray}
Substituting asymptotic solutions for our field fluctuations into \eqref{Eq:boundarysourceidentifications} we find the following relations
    \begin{eqnarray}
     \label{Eq:boundaryidentifications}
     \langle J^{i} \rangle &=& (F^{-1})^{ij} \left( n_{f} E_{j} + ( s_{f} T - P_{l} - m B) \frac{\partial_{j} T}{T} - \delta s_{j} \right) \; , \qquad \bar{E}_{i} = \frac{\partial_{i} T}{n_{f} T} \; . \qquad 
    \end{eqnarray}
Consequently in the spontaneous case one can immediately read off all DC values of the transport coefficients involving the electric charge current i.e.~
    \begin{eqnarray}
     \sigma_{(\mathrm{H})}(0) = - n_{f} \; , \; \;
     \alpha_{(\mathrm{H})}(0) = - \left( s_{f} T - m B - P_{l} \right) \; , \; \;
     \gamma_{(\mathrm{H})}(0) = - 1 \; , 
    \end{eqnarray}
with all other conductivities involving the current operator being zero. We can derive these without reference to the near horizon values of the bulk current because they are the transport coefficients dictated by symmetry.}

{\ The other DC observables in the spontaneous case can be found by matching the near horizon and boundary values of the conserved radial heat currents. Unfixed thus far are the expectation value of the boundary stress tensor and the values of the sliding mode coefficients $\delta V_{i}$. Employing our expressions \eqref{Eq:boundaryidentifications}, one can write the bulk heat currents $\delta \mathcal{Q}_{i}$ in terms of $E_{i}$, $\partial_{i} T/T$ and $\delta s_{i}$. The resultant thermal conductivities obtained from this bulk current are then:
    \begin{eqnarray}
      \kappa_{(\mathrm{L})}(0) = \frac{1}{T} \left( \frac{Z_{\mathrm{h}} \left(s_{f} T- P_{l}\right)^2}{n_{f}^2 + B^2 Z_{\mathrm{h}}^2} + \frac{4 \pi  P_{l}^2}{s_{f} Y_{\mathrm{h}}} \right) \; , \; \; 
      \kappa_{(\mathrm{H})}(0) = - \frac{n_{f} \left(s_{f} T - P_{l} \right)^2}{T \left(n_{f}^2 + B^2 Z_{\mathrm{h}}^2 \right)} - \frac{M_{Q}}{n_{f} T} \; , \qquad
    \end{eqnarray}
and the thermal-Goldstone zero frequency terms are 
    \begin{eqnarray}
     \theta_{(\mathrm{L})}(0) = \frac{4 \pi I_{Y}}{s Y_{\mathrm{h}}} - \frac{( s_{f} T - P_{l} ) Z_{h}}{n_{f}^2 + B^2 Z_{\mathrm{h}}^2} \; , \qquad
     \theta_{(\mathrm{H})}(0) = \frac{n_{f} ( s_{f} T - P_{l} ) Z_{h}}{n_{f}^2 + B^2 Z_{\mathrm{h}}^2} \; . 
    \end{eqnarray}
Meanwhile, the DC term corresponding to the Goldstone-Goldstone correlator can be obtained from \eqref{eqn:configuration_equation}. To do this we work in equilibrium, vary both sides of equation \eqref{eqn:configuration_equation} with respect to $\delta s_{i}$ and then take the limit of $\omega \rightarrow 0$. In this limit we can identify $X^{ij}(0)$ with the residue of $\langle O_{i} O_{j} \rangle$ at $\omega=0$, see \eqref{Eq:ACSpontaneousconductivities}. Consequently we find
    \begin{eqnarray}
      X^{ij}(0) = - \frac{\delta V^{i}}{\delta s_{j}} \; , \qquad
    \end{eqnarray}
 As we have already determined the sliding mode $\delta V_{i}$ in terms of the boundary sources (the expression is too long to include here) we readily find
    \begin{eqnarray}
      X_{(\mathrm{L})}(0) = - \left( \frac{4 \pi}{k^2 s_{f} Y_{\mathrm{h}}} + \frac{Z_{\mathrm{h}}}{n_{f}^2 + B^2 Z_{\mathrm{h}}^2} \right) \; , \qquad
      X_{(\mathrm{H})}(0) = \frac{n_{f}}{(n_{f}^2 + Z_{\mathrm{h}}^2 B^2)} \; . \qquad
    \end{eqnarray}
}

\subsubsection{Explicit case}

{\ The expressions for the bulk electric charge currents \eqref{Eq:BulkChargeCurrents} are unchanged between the explicit and spontaneous case. However the identifications of the constants $p_{x}^{(0)}$ and $p_{y}^{(0)}$ of \eqref{Eq:p1p2rsols} are different
    \begin{subequations}
    \label{Eq:p1p2identificationexplicit}
    \begin{eqnarray}
        p_{x}^{(0)} &=& \left( s_{f} T - m B - k^2 I_{Y} \right) \bar{E}_{x} - \frac{B}{n_{f}} \langle J^{y} \rangle - \frac{\langle O^{x} \rangle}{n_{f}} \; , \\
        p_{y}^{(0)} &=& \left( s_{f} T - m B - k^2 I_{Y} \right) \bar{E}_{y} + \frac{B}{n_{f}} \langle J^{x} \rangle - \frac{\langle O^{y} \rangle}{n_{f}} \; , \\
        \label{Eq:p1p2identificationexplicitvevs}
        \langle O^{i} \rangle &:=& k \lambda^2 \chi_{i}'(0) \; .
    \end{eqnarray}
    \end{subequations}
Consequently, in the explicit case, the expressions for the boundary current \eqref{Eq:boundaryidentifications} are modified to
    \begin{eqnarray}
     \label{Eq:boundaryidentificationsexplicit}
     \langle J^{i} \rangle &=& (F^{-1})^{ij} \left( n_{f} E_{j} + ( s_{f} T - k^2 I_{Y} - m B) \frac{\partial_{j} T}{T} +  \langle O_{j} \rangle \right) \; , \qquad 
    \end{eqnarray}
where again we have employed \eqref{Eq:boundarysourceidentifications}.}

{\ The bulk (radially conserved) heat currents in the explicit case are
    \begin{eqnarray}
        \delta \mathcal{Q}_{x}(r) &=& - f(r) \left( \frac{h_{x}(r)}{r^2} \right)' + \left( \frac{f(r)}{r^2} \right)' h_{x}(r) \nonumber \\
        &\;& + \left( a(r) - \frac{B^2}{n_{f}} I_{Z}(r) \right) \langle J^{x} \rangle
        + \frac{B}{ n_{f}} \langle O_{y} \rangle I_{Z}(r) \qquad \nonumber \\
        &\;& + B \bar{E}_{y} \left( M_{Q}(r) - (s_{f} T + n_{f} a(r) - m B - k^2 I_{Y} ) I_{Z}(r) \right) \; , \qquad \\
        \delta \mathcal{Q}_{y}(r) &=& - f(r) \left( \frac{h_{y}(r)}{r^2} \right)' + \left( \frac{f(r)}{r^2} \right)' h_{y}(r) \nonumber \\
        &\;& + \left( a(r) - \frac{B^2}{n_{f}} I_{Z}(r) \right) \langle J^{y} \rangle
        - \frac{B}{ n_{f}} \langle O_{x} \rangle I_{Z}(r) \qquad \nonumber \\
        &\;& - B \bar{E}_{x} \left( M_{Q}(r) - (s_{f} T + n_{f} a(r) - m B - k^2 I_{Y} ) I_{Z}(r) \right) \; , 
    \end{eqnarray}
These differ from the spontaneous case through the dropping of terms dependent on the sliding modes $\delta V_{i}$ and the replacement of $\delta s_{i}$ by terms proportional to $\langle O_{i}\rangle$.}

{\ Once more, we use our identifications in \eqref{Eq:boundarysourceidentifications}, and the matching of the bulk currents at boundary and horizon, to express the subleading term in the boundary expansion of the bulk graviton and $ \langle O_{i} \rangle$ in terms of the boundary electric field $E_{i}$ and the temperature gradient $\partial_{i} T/T$ - there is no explicit axion source term in our expressions to worry about. Doing this fixes $\langle O_{i} \rangle$ appearing in \eqref{Eq:boundaryidentificationsexplicit} in terms of the boundary fluctuations of temperature and electric field. Consequently we can read off the various DC thermo-electric conductivities. For example, the electric charge conductivities are
    \begin{eqnarray}
     \sigma_{(\mathrm{L})}(0) &=& \frac{k^2 s_{f} Y_{\mathrm{h}} \left( k^2 s_{f} Y_{\mathrm{h}} Z_{\mathrm{h}} + 4 \pi ( n_{f}^2 + B^2 Z_{\mathrm{h}}^2 ) \right)}{(4 \pi n_{f} B)^2 + \left( k^2 s_{f} Y_{\mathrm{h}} + 4 \pi Z_{\mathrm{h}} B^2 \right)^2} \; , \\
     \sigma_{(\mathrm{H})}(0) &=& - \frac{8 \pi n_{f} \left( k^2 s_{f} Y_{\mathrm{h}} Z_{\mathrm{h}} + 2 \pi ( n_{f}^2 + B^2 Z_{\mathrm{h}}^2 ) \right)}{(4 \pi n_{f} B)^2 + \left( k^2 s_{f} Y_{\mathrm{h}} + 4 \pi Z_{\mathrm{h}} B^2 \right)^2} \; . 
    \end{eqnarray}
\captionsetup[subfigure]{labelformat=empty}
\begin{figure}
	\centering
	\begin{subfigure}{.31\textwidth}
		\centering
		\includegraphics[width=\linewidth]{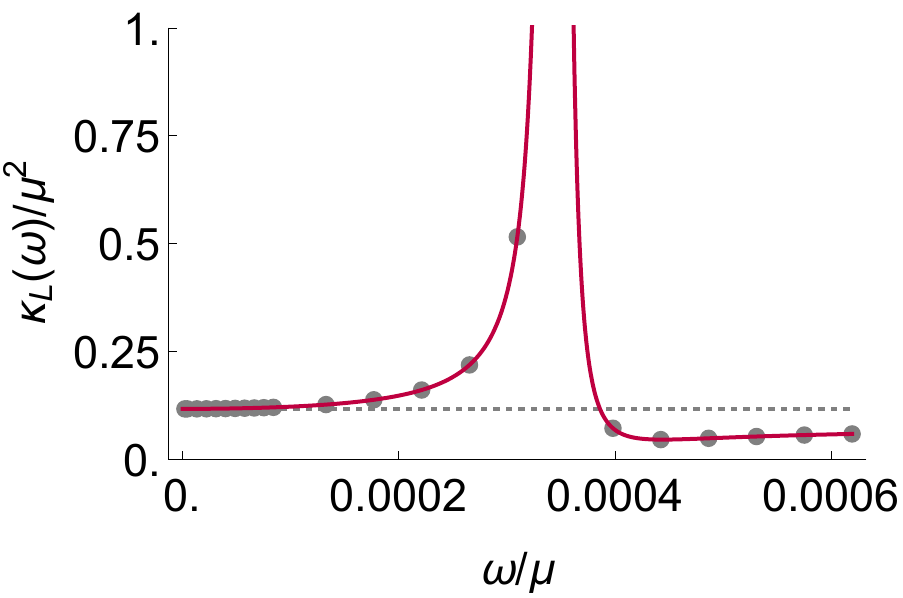}
	\end{subfigure} \hspace{.01\textwidth}
	\begin{subfigure}{.31\textwidth}
		\centering
		\includegraphics[width=\linewidth]{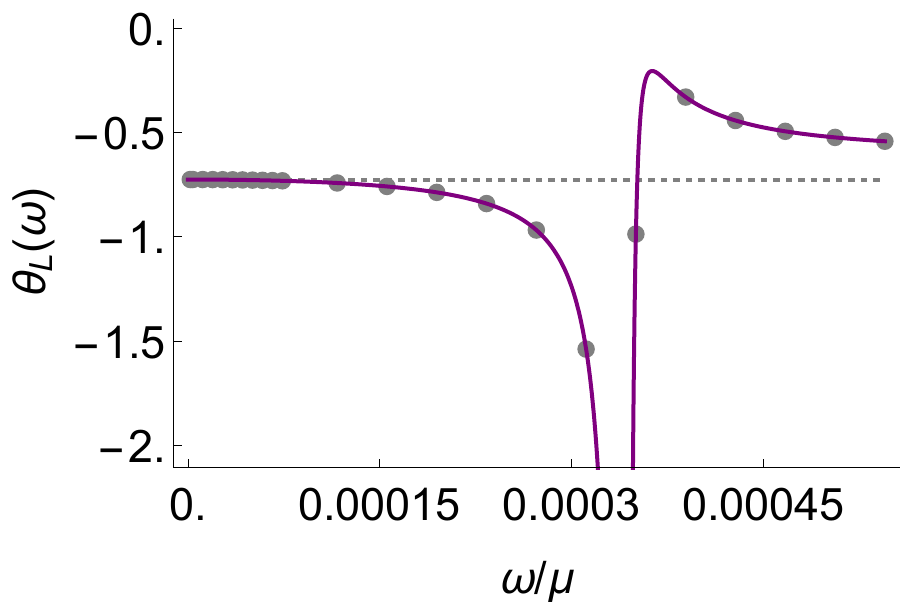}
	\end{subfigure} \hspace{.01\textwidth}
	\begin{subfigure}{.31\textwidth}
		\centering
		\includegraphics[width=\linewidth]{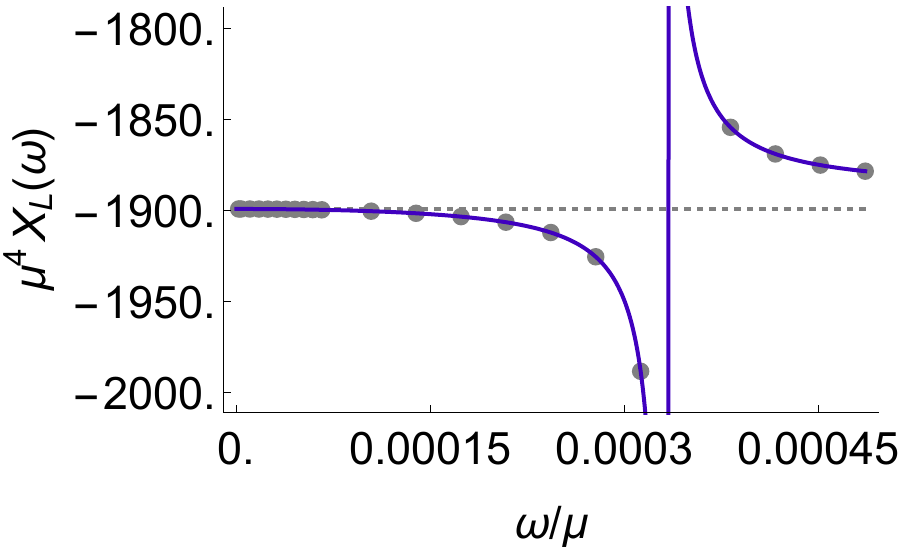}
	\end{subfigure} \hfill \\
	\begin{subfigure}{.31\textwidth}
		\centering
		\includegraphics[width=\linewidth]{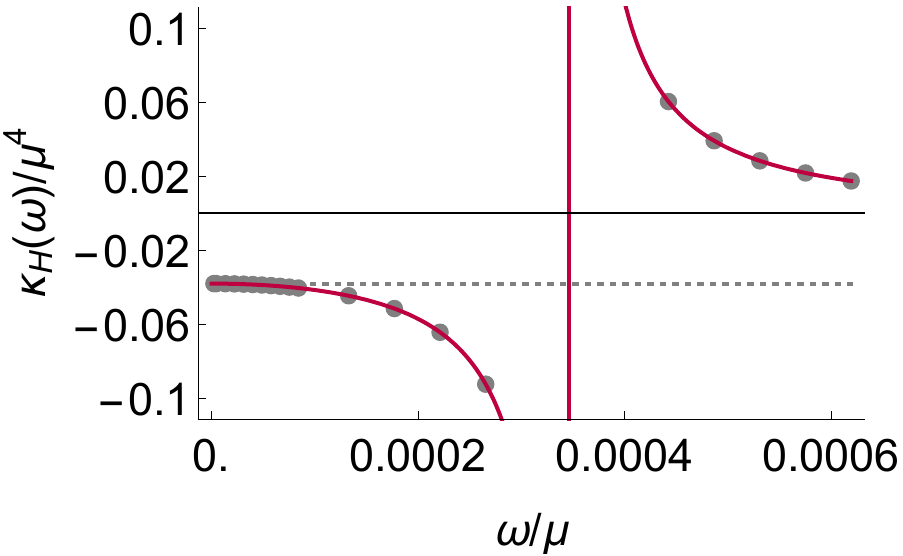}
	\end{subfigure}  \hspace{.01\textwidth}
	\begin{subfigure}{.31\textwidth}
		\centering
		\includegraphics[width=\linewidth]{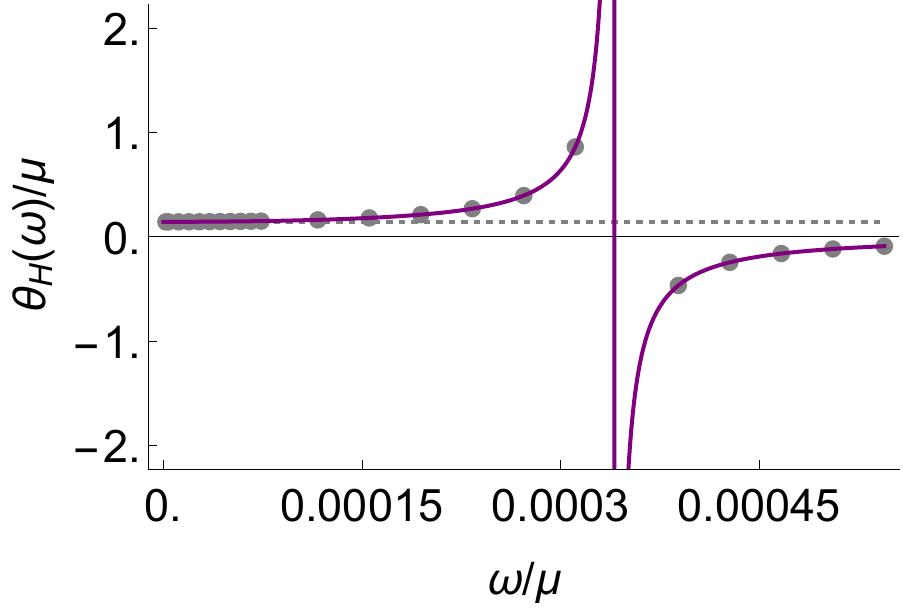}
	\end{subfigure}  \hspace{.01\textwidth}
	\begin{subfigure}{.31\textwidth}
		\centering
		\includegraphics[width=\linewidth]{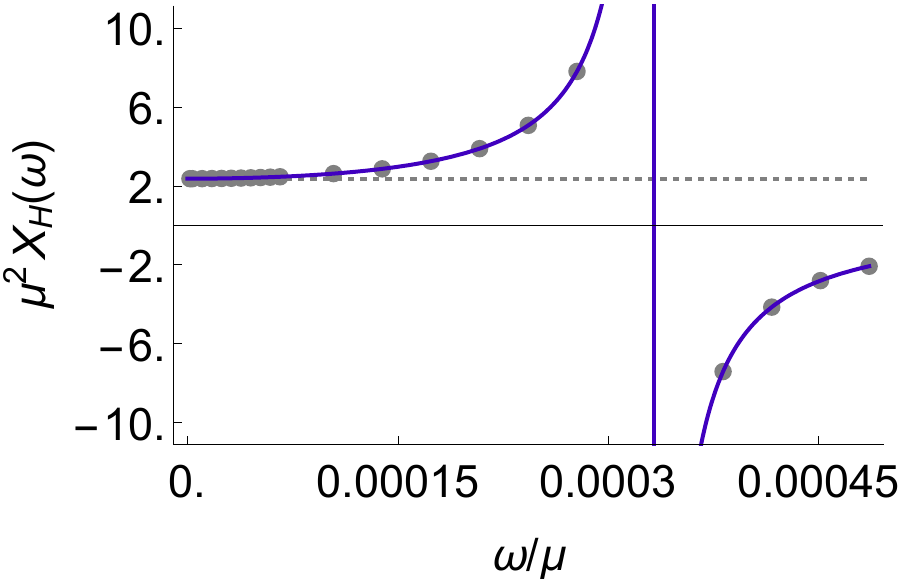} 
	\end{subfigure} \hfill
	\caption{Spontaneous case AC correlators at $k/\mu=10^{-1}$.
		Grey dots are numerical data, solid lines are our analytic hydrodynamic expressions and the dashed grey line is the DC value of the coefficient. \textbf{Left column:} The thermal conductivities ($\kappa(\omega)$) at $T/\mu = 0.06$ and $B/\mu^2 \approx 4.4 \times 10^{-4}$. \textbf{Central column:} The heat-Goldstone correlators ($\theta(\omega)$) at $T/\mu=0.04$ and $B/\mu^2 \approx 3.9 \times 10^{-4}$. \textbf{Right column:} The Goldstone-Goldstone correlators ($X(\omega)$) at $T/\mu=0.02$ and $B/\mu^2 \approx 3.5 \times 10^{-4}$.}
	\label{fig:spontaneousAreanACconductivities}
\end{figure} 

The rest of the expressions are relegated to appendix \ref{appendix:formulae}. Our expressions involving the conserved currents agree with those computed in \cite{Blake:2015ina} and we have also determined the zero frequency limits for correlators involving the scalars. Finally, from the DC limit of the electric-axion correlator one can identify
    \begin{eqnarray}
     \zeta_{(\mathrm{L})}(0) &=& \frac{ n_{f} (k^2 s_{f} Y_{\mathrm{h}})^2}{(4 \pi n_{f} B)^2 + \left(k^2 s_{f} Y_{\mathrm{h}} + 4 \pi B^2 Z_{\mathrm{h}}\right)^2} \; , \qquad  \\
     \zeta_{(\mathrm{H})}(0) &=& \frac{ k^2 s_{f} B Y_{\mathrm{h}} \left(k^2 s_{f} Y_{\mathrm{h}} Z_{\mathrm{h}} + 4 \pi \left(B^2 Z_{\mathrm{h}}^2 + n_{f}^2 \right)\right)}
                                 {(4 \pi n_{f} B)^2 + \left(k^2 s_{f} Y_{\mathrm{h}} + 4 \pi B^2 Z_{\mathrm{h}}\right)^2} \; 
    \end{eqnarray}
 The remaining DC terms in the explicit case are listed in appendix \ref{appendix:formulae}.}

\subsection{AC correlators}

{\ With exact expressions for the DC values of the various correlators we can now employ our hydrodynamic expressions for the correlators and compare to the equivalent quantities obtained from holography. Naturally many of our observations in this section will be model dependent at the quantitative level; however certain features we expect to hold in general models. Moreover, we can put our hydrodynamic theory to a precision test.}

\subsubsection{Spontaneous case}

{\ There are twelve potential AC correlators to display including the transport coefficients, the current-Goldstone correlators and the Goldstone-Goldstone correlators. Given the difficulties in the past of matching the thermal DC conductivities to hydrodynamics we shall choose to display these and not the other conductivities as they tend to be quite robust to small errors. We shall also show the AC thermal-Goldstone correlator and the Goldstone-Goldstone correlators with a magnetic field as these types of correlators are novel in the literature.}

\captionsetup[subfigure]{labelformat=empty}
\begin{figure}
        \centering
        \includegraphics[width=0.47\textwidth]{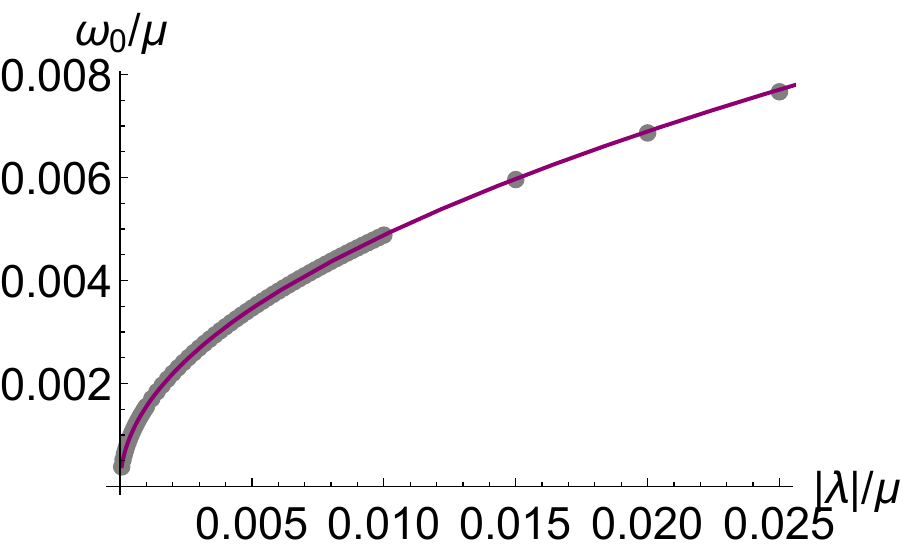}
 \caption{The pinning frequency against $\lambda/\mu$ at $k/\mu = 0.1$, $B/\mu^2 = 10^{-2}$ and $T/\mu = 5 \times 10^{-2}$. The solid purple line is a best fit to the data points with an expression proportional to $\sqrt{|\lambda|/\mu}$.}
    \label{fig:omega0flowwithlambda}
\end{figure}

{\ As can be seen in fig.~\ref{fig:spontaneousAreanACconductivities} the matching between our analytic hydrodynamic expressions and the holographic model is excellent over a wide range of parameters. In fact, it is somewhat surprising that they work so well down to rather low temperatures and relatively high magnetic fields. There exists one peak at $\omega>0$ in the electric, thermo-electric and thermal conductivities corresponding to the cyclotron mode.}

{\ Somewhat new to the literature, but perhaps not unexpected, is the smoothing of the low frequency Goldstone-Goldstone correlators. It was observed in previous works \cite{Amoretti_2019} that these correlators have a double pole in frequency located at $\omega=0$. Taking the zero magnetic field limit of our longitudinal expression for the Goldstone-Goldstone correlator one again finds this double pole emerging. For finite $B$ however one of the degenerate poles is displaced and becomes the cyclotron modes; leaving an isolated pole at $\omega =0$. This can be seen from the lack of any $\omega \rightarrow 0$ divergence in $X_{(\mathrm{L})}(\omega)$ and $X_{(\mathrm{H})}(\omega)$ as displayed in fig.~\ref{fig:spontaneousAreanACconductivities}.}

\subsubsection{Explicit case}

{\ Hydrodynamics and the DC conductivities almost fix the transport coefficients appearing in our hydrodynamic expressions, \eqref{Eq:ExplicitACtransportcoeffs}, completely. There remains a single parameter that must be determined numerically: $\omega_{0}$. This is the pinning frequency of the phonon-like mode. There are a couple of methods by which this may be determined, and we have tested that both are consistent. Firstly, one may examine the quasinormal modes of the theory and solve for $\omega_{0}$ using their position in the complex plane. Alternatively, one may take any of the correlators at low frequency (so that we are in the hydrodynamic regime) and request that the analytic expression match the numerically determined one. Doing so allows one to solve for the pinning frequency.}

\captionsetup[subfigure]{labelformat=empty}
\begin{figure}
\begin{subfigure}{.47\textwidth}
  \centering
  \includegraphics[width=\linewidth]{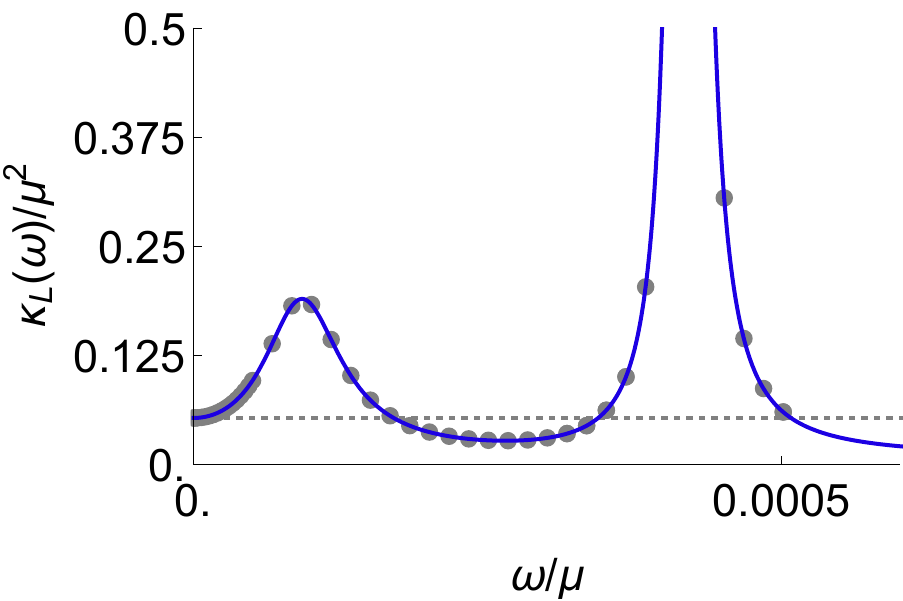}
\end{subfigure} \hfill %
\begin{subfigure}{.47\textwidth}
  \centering
  \includegraphics[width=\linewidth]{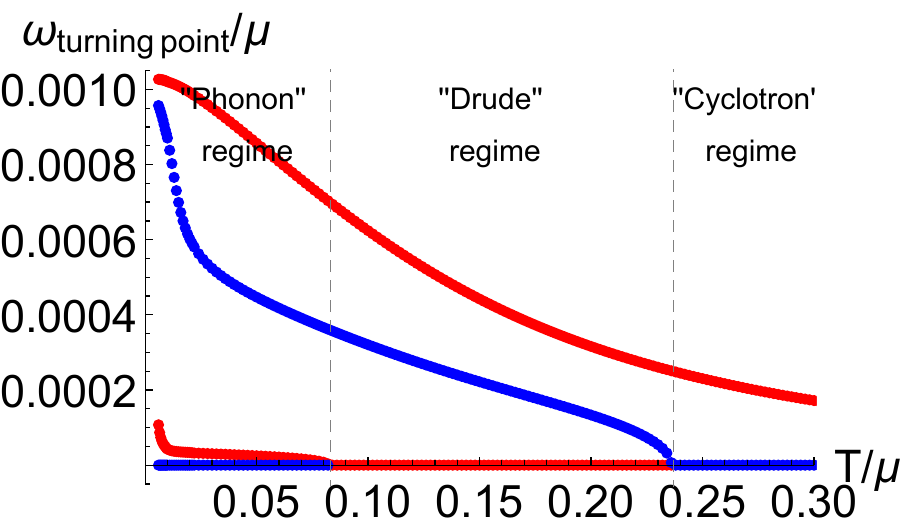}
\end{subfigure}
\caption{Conductivities in the (pseudo-) explicit breaking regime.
		\textbf{Left:} The AC longitudinal thermal conductivity at $\lambda/\mu= - 10^{-5}$, $k/\mu =0.1$, $B/\mu^2 \approx 3 \times 10^{-4}$ and $T/\mu = 10^{-2}$. Notice 
		the two peaks, both displaced from $\omega = 0$. \textbf{Right:} The frequency of the maxima (red) and minima (blue) in the hydrodynamic longitudinal electric charge conductivity as a function of temperature at $\lambda/\mu=-10^{-5}$, $k/\mu=0.1$ and $B/\mu^2 = 10^{-3}$. At the lowest temperatures we have two peaks in the $\omega>0$ half-line and also a minimum at $\omega=0$. As the temperature increases the two pseudo-Goldstone modes join ($T/\mu \approx 0.083$) to become a single Drude-like peak at zero frequency. This zero frequency peak eventually drops out of the correlator, becoming a trough, at $T/\mu \approx 0.237$. }
\label{fig:Twopeaksexplicit}
\end{figure}

{\ Regarding the pinning frequency, for a range of $|\lambda/\mu| \in (10^{-5},10^{-2})$ we have found that the pinning frequency $\omega_{0}$ is proportional to $\sqrt{|\lambda|/\mu}$. This is in accordance with the behavior found in the same holographic model at zero magnetic field in \cite{Amoretti:2018tzw} and with more general quantum field theory arguments, as explained in \cite{Amoretti:2016bxs}. We display the flow of $\omega_{0}$ with $\lambda$ for a particular choice of temperature and magnetic field in fig.~\ref{fig:omega0flowwithlambda}. For increasing temperature the pinning frequency gets progressively smaller.}

{\ Now let us consider the behaviour of the suite of thermo-electric AC conductivities. In figures \ref{fig:Twopeaksexplicit} and \ref{fig:conductivities} we plot the electric and thermal conductivities showing an excellent agreement between our hydrodynamic expressions \eqref{Eq:ExplicitACconductivities} and the exact holographic data.
Notice that in the explicit case it is possible that these thermo-electric correlators have two peaks on the $\omega >0$ half-line, both displaced from $\omega=0$ to some finite value of $\omega$; i.e.~the point $\omega=0$ is a minimum.
An example of this phenomenon is displayed in the left hand plot of fig.~\ref{fig:Twopeaksexplicit} for the longitudinal thermal correlator. On the right hand side of the same figure we show the flow of the maxima and minima of the analytic hydrodynamic longitudinal conductivity as a function of $T/\mu$ for a particular choice of $\lambda/\mu$ and $B/\mu^2$. 
One can identify a  low temperature `phonon regime' where two peaks at nonzero
frequency are observed. At intermediate temperatures the correlator displays a Drude-like peak at the origin and a cyclotron peak at finite $\omega$. We denote
this temperature range as `Drude regime'.
These two peaks,
and the underlying quasinormal modes,
can be qualitatively interpreted
as the magnetophonon and mangnetoplasmon resonances expected in the
hydrodynamic regime of a weakly-pinned Wigner crystal~\cite{Delacretaz:2019wzh}.
At high temperatures
only a cyclotron peak at finite $\omega$ is observed and we expect the correlator to
be well described by magnetohydrodynamics with a decay rate (i.e.~see \cite{Sachdev} for further discussion).
The two inflection points marking the transition between the three regimes
occur when the following conditions are satisfied
\begin{eqnarray}
\zeta_{(\mathrm{L})}'(0) = 0 \, , \qquad
\zeta_{(\mathrm{L})}''(0) = 2 \left( 3 \mu^2 \sigma_{\mathrm{L}}(0) + 4 \mu \alpha_{(\mathrm{L})}(0) + \kappa_{(\mathrm{L})}(0) \right) \,,
\end{eqnarray}
which can be obtained from the Ward identity by requiring $\omega=0$ to be an inflection point.}

{\ In the Drude-like regime the peak at nonzero frequency is associated with the cyclotron mode and the corresponding pair of quasinormal modes.
The peak at $\omega=0$ is however a little unusual in that it is not associated with a single (imaginary) quasinormal mode, but instead with two complex modes. While we have termed this the `Drude regime' on account of the single peak at small frequency, one must be careful in interpreting this since, as explained in details in \cite{amoretti:hydrodynamicmagnetotransport}, it has nothing to do with any explicit coherent momentum decay rate in the hydrodynamic theory. If one looks at hydrodynamics in an external magnetic field with a non-zero momentum decay rate tensor $\Gamma^{ij}$, and no translation breaking scalars, one finds only two quasinormal modes in the diffusive sector which can be identified as displaced cyclotron modes. In particular there is no Drude-like peak in such a system.}

\captionsetup[subfigure]{labelformat=empty}
\begin{figure}
    \begin{subfigure}{.47\textwidth}
        \centering
        \includegraphics[width=\linewidth]{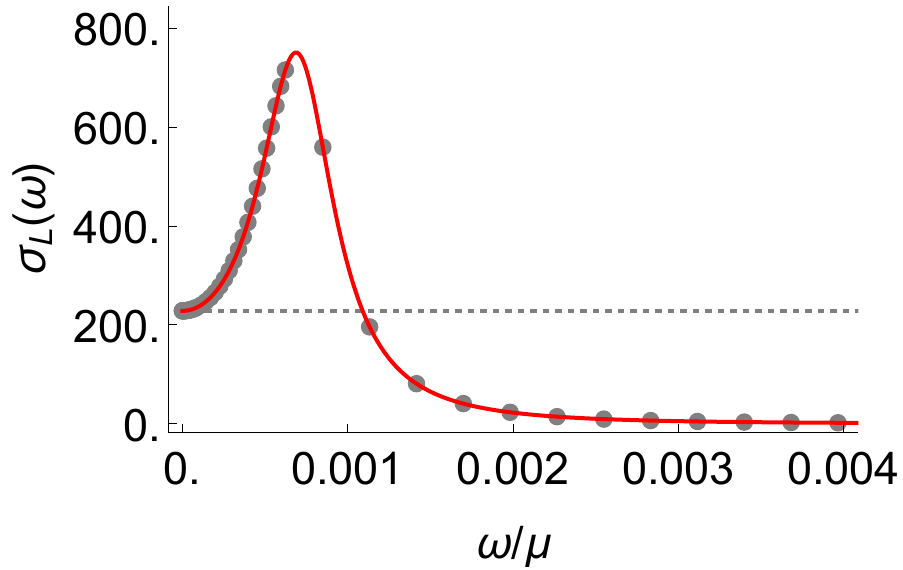}
    \end{subfigure} \hfill %
    \begin{subfigure}{.47\textwidth}
        \centering
        \includegraphics[width=\linewidth]{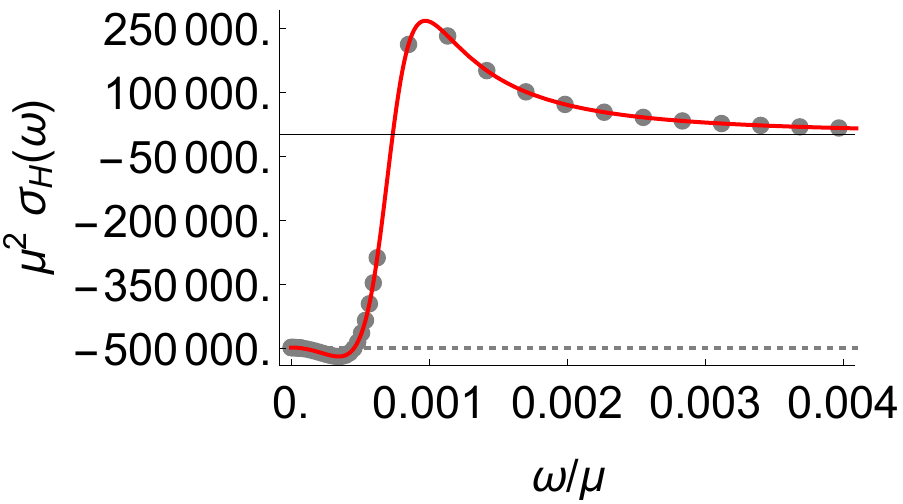}
    \end{subfigure} \\
    \begin{subfigure}{.47\textwidth}
        \centering
        \includegraphics[width=\linewidth]{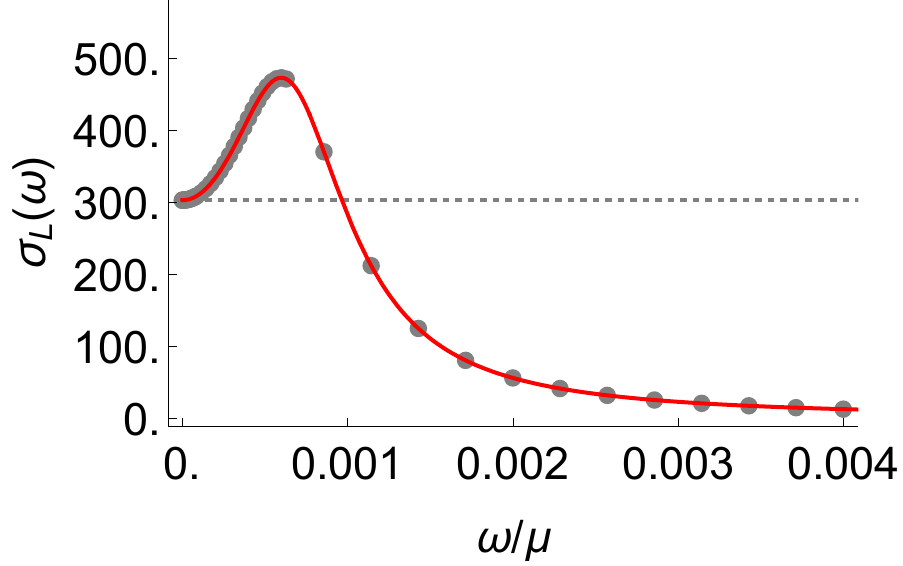}
    \end{subfigure} \hfill %
    \begin{subfigure}{.47\textwidth}
        \centering
        \includegraphics[width=\linewidth]{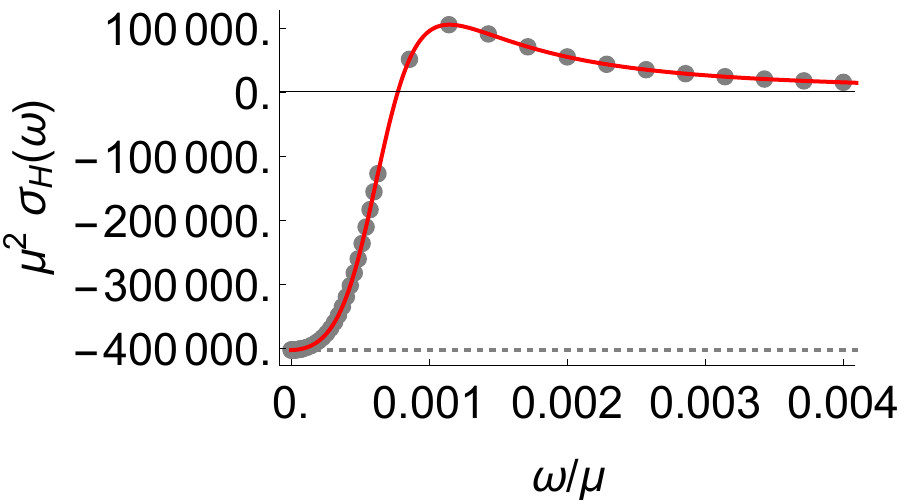}
    \end{subfigure}
    \caption{AC electric conductivities at $B/\mu^2=10^{-3}$, $T/\mu=10^{-1}$ and $k/\mu=10^{-1}$. Red lines are the analytic hydrodynamic expressions while grey dots are numerical data. \textbf{Left:} The longitudinal conductivities in two regimes - the pseudo-spontaneous (top) where $\lambda \mu/\phi_{v} \approx 0.004$ and a strongly explicit regime (bottom) where $\lambda \mu/\phi_{v} \approx 0.95$. In both cases our hydrodynamic expressions closely match the data. \textbf{Right:} The Hall conductivities in the same regimes.}
    \label{fig:conductivities}
\end{figure}

\subsection{On the spurious pole}

{\ Our final observation concerns the number of poles implied by our hydrodynamic expressions. Curiously, the formalism of \cite{Armas:2019sbe,Armas:2020bmo} predicts the existence of an additional gapped pole with respect to the hydrodynamic approach of \cite{amoretti:hydrodynamicmagnetotransport}. The existence of this additional pole is related to the presence of the lattice pressure $P_l$ as one can see from the frequency dependent term in \eqref{Gammafreq}, which gives rise to an extra zero in the denominator of the correlators \eqref{denominatore}, not present in hydrodynamic approach of \cite{Delacretaz:2017zxd,Delacretaz:2019wzh,amoretti:hydrodynamicmagnetotransport}. For the systems we have investigated, this extra pole has always a very large imaginary part (which we checked numerically) and its effect on the diffusive correlators can be ignored.}

\captionsetup[subfigure]{labelformat=empty}
\begin{figure}
  \centering
  \includegraphics[width=0.47\linewidth]{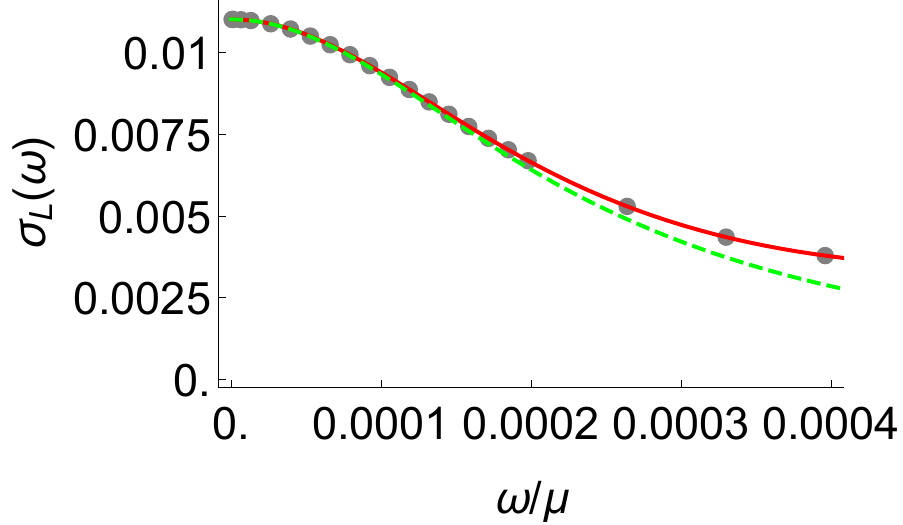}
\caption{The longitudinal electric charge conductivity against frequency at small frequencies in the Drude-like regime with $\lambda/\mu = -10^{-5}$, $k/\mu=10^{-1}$, $T/\mu = 0.3$ and $B/\mu^2 \approx 0.066$.
	The solid red line corresponds to the hydrodynamic expression, the grey dots are
	numerical data, and the dashed Green line shows a pure Drude-like approximation $\sim 1/ (\omega - i \Gamma)$.}
\label{fig:Drudeemergence}
\end{figure}

{\ In the papers where this formalism was developed a small frequency expansion was taken to eliminate this additional pole from the diffusive sector \cite{privatecomm}. In our systems, we could not take a low frequency expansion without washing out all of our poles as the external magnetic field gaps the system. Instead, when comparing our expressions with data, we checked that this additional ``spurious'' pole must reside deep in the complex plane according to our hydrodynamic expressions and as such could be ignored.}

{\ We did check to determine whether there was a hint that this spurious pole existed within the system by examining the numerical correlators at complex frequency around the position predicted by our hydrodynamic expressions. Thus far we have found no evidence of its presence in the diffusive part of the spectrum. In fact, we found that other quasinormal modes become relevant before any hint of this additional pole appears.}

In order to check that this additional pole is not an artifact of the frame choice, we computed the Green functions for the spontaneous case using the method described in Sec~\ref{sec:2}, at zero magnetic field and in two distinct frames, and we found the same results as for the Landau frame. In particular we considered a pseudo-Eckart frame (eliminating all first derivative terms from the electric charge current \eqref{eqn:constitutive_relations} except for the $\gamma$ term) and the true Eckart frame where $J^{\mu} = q u^{\mu}$ to all orders in derivatives. Notice that the $\gamma$ term in \eqref{eqn:constitutive_relations} is naively order zero in derivatives until one substitutes for $u^\mu e^I_\mu$ using the configuration equation.

\section{Conclusions}
\label{sec:conclusions}
In this paper we have provided a complete hydrodynamic description of holographic Q-lattice models which present a spontaneous or a pseudo-spontaneous breaking of translations in the presence of an external magnetic field. To take into account the presence of a non-trivial lattice pressure term $P_l$ in the thermodynamics of these holographic models, we have generalized the hydrodynamic approach of \cite{Armas:2019sbe,Armas:2020bmo} in order to include both a small mass for the Goldstone boson related to translation symmetry breaking and an external magnetic field. Moreover, using the method of \cite{amoretti:magnetothermaltransporta,amoretti:hydrodynamicmagnetotransport}, we have been able to express all the hydrodynamic AC correlators in terms of their DC values and the pinning frequency. Since the DC holographic thermo-electric conductivities have a closed analytic form in terms of horizon data, combining the hydrodynamic result with holography we have  provided an analytic form for the holographic correlators in terms of the horizon data of the model and one undetermined quantity, the pinning frequency, which we have obtained numerically. The correlators computed in this way match excellently the holographic numerical result, and the behavior of the pinning frequency agrees with the one reported previously for the same model in the absence of an external magnetic field \cite{Amoretti:2018tzw}. 
Finally, the identification of a regime where the AC correlators feature a deep IR peak that can be identified with  the magnetophonon collective mode opens the way
	for a further exploration of the holographic Q-lattice models as  avatars of 
	strongly coupled electronic phases of matter.


\section*{Acknowledgments}
We would like to thank Jay Armas and Akash Jain for private communications about the hydrodynamic model described in this paper. We also thank Blaise Gout\'eraux for a careful reading of a previous version of the present paper. The project has been partially supported by the INFN Scientific Initiative SFT: ``Statistical Field Theory, Low-Dimensional Systems, Integrable Models and Applications''. D. A. is supported by the `Atracci\'on de Talento' programme
(2017-T1/TIC-5258, Comunidad de Madrid) and through the grants SEV-2016-0597 and PGC2018-095976-B-C21.
\appendix

\section{Explicit DC transport coefficients}
\label{appendix:formulae}

{\ In this appendix we list the expressions for the DC transport coefficients that follow from the computations in section~\ref{ssec:holoDCxp} with
    \begin{eqnarray}
     \vartheta^{ij}(\omega) = \frac{1}{i\omega} \langle Q^{i} O^{j} \rangle \; . 
    \end{eqnarray}
}

\setlength\LTleft{-0.8cm}
\setlength\LTright{0pt plus 1fill minus 1fill}
\def\arraystretch{1.5}
    \begin{longtable}{|l|c|c|}
        \hline \rowcolor{lightgray} 
        & \multicolumn{2}{c|}{\textbf{Explicit}} \\ \hline
        & (L) & (H) \\ \hline
        $\hat{\sigma}(0)$ &  $\frac{k^2 s_{f} Y_{\mathrm{h}} \left( k^2 s_{f} Y_{\mathrm{h}} Z_{\mathrm{h}} + 4 \pi ( n_{f}^2 + B^2 Z_{\mathrm{h}}^2 ) \right)}{(4 \pi n_{f} B)^2 + \left( k^2 s_{f} Y_{\mathrm{h}} + 4 \pi Z_{\mathrm{h}} B^2 \right)^2} $ & $- \frac{8 \pi n_{f} \left( k^2 s_{f} Y_{\mathrm{h}} Z_{\mathrm{h}} + 2 \pi ( n_{f}^2 + B^2 Z_{\mathrm{h}}^2 ) \right)}{(4 \pi n_{f} B)^2 + \left( k^2 s_{f} Y_{\mathrm{h}} + 4 \pi Z_{\mathrm{h}} B^2 \right)^2}$ \\ \hline 
        $\hat{\alpha}(0)$ & $ \frac{4 \pi k^2 s_{f}^2 T n_{f} Y_{\mathrm{h}}}{(4 \pi n_{f} B)^2 + \left( k^2 s_{f} Y_{\mathrm{h}} + 4 \pi Z_{\mathrm{h}} B^2 \right)^2}$ & $ \frac{(4 \pi n_{f})^2 (B m - s_{f} T) + (k^2 s_{f} Y_{\mathrm{h}} + 4 \pi B^2 Z_{\mathrm{h}}) (k^2 m s_{f} Y_{\mathrm{h}} - 4 \pi B Z_{\mathrm{h}}(s_{f} T - m B))}{(4 \pi n_{f} B)^2 + \left( k^2 s_{f} Y_{\mathrm{h}} + 4 \pi Z_{\mathrm{h}} B^2 \right)^2}$ \\ \hline
        $\hat{\kappa}(0)$ & $ \frac{4 \pi (s_{f} T)^2 \left( k^2 s_{f} Y_{\mathrm{h}}+ 4 \pi B^2 Z_{\mathrm{h}} ) \right)}{(4 \pi n_{f} B)^2 + \left( k^2 s_{f} Y_{\mathrm{h}} + 4 \pi Z_{\mathrm{h}} B^2 \right)^2}$ & $ - \frac{n_{f} (4 \pi s_{f} T)^2}{(4 \pi n_{f} B)^2 + \left( k^2 s_{f} Y_{\mathrm{h}} + 4 \pi Z_{\mathrm{h}} B^2 \right)^2} - \frac{M_{Q}}{n_{f} B}$ \\ \hline
        $\hat\varpi(0)$ & $ \frac{n_{f} (k^2 s_{f} Y_{\mathrm{h}})^2}{(4 \pi n_{f} B)^2 + \left( k^2 s_{f} Y_{\mathrm{h}} + 4 \pi Z_{\mathrm{h}} B^2 \right)^2}$ & $ - \frac{k^2 s_{f} Y_{\mathrm{h}} \left( k^2 s_{f} Y_{\mathrm{h}} Z_{\mathrm{h}} + 4 \pi ( n_{f}^2 + B^2 Z_{\mathrm{h}}^{2}) \right)}{(4 \pi n_{f} B)^2 + \left( k^2 s_{f} Y_{\mathrm{h}} + 4 \pi Z_{\mathrm{h}} B^2 \right)^2}$ \\ \hline
        $\hat\vartheta(0)$ & $\frac{ s_{f}^2 T Y_{\mathrm{h}} \left( k^2 s_{f} Y_{\mathrm{h}} + 4 \pi B^2 Z_{\mathrm{h}} \right)}{(4 \pi n_{f} B)^2 + \left( k^2 s_{f} Y_{\mathrm{h}} + 4 \pi Z_{\mathrm{h}} B^2 \right)^2} - k^{2} I_{Y}$ & $- \frac{4 \pi k^2 s_{f}^2 T n_{f} Y_{\mathrm{h}}}{(4 \pi n_{f} B)^2 + \left( k^2 s_{f} Y_{\mathrm{h}} + 4 \pi Z_{\mathrm{h}} B^2 \right)^2}$ \\ \hline
        $\hat\zeta(0)$ & $-\frac{4 \pi k^2 s_{f}^2 T n_{f} Y_{\mathrm{h}}}{(4 \pi n_{f} B)^2 + \left( k^2 s_{f} Y_{\mathrm{h}} + 4 \pi Z_{\mathrm{h}} B^2 \right)^2}$ & $-\frac{ B^2 k^2 s_{f} Y_{\mathrm{h}} \left( k^2 s_{f} Y_{\mathrm{h}} Z_{\mathrm{h}} + 4 \pi (n_{f}^2 + B^2 Z_{\mathrm{h}}^{2}) \right)}{(4 \pi n_{f} B)^2 + \left( k^2 s_{f} Y_{\mathrm{h}} + 4 \pi Z_{\mathrm{h}} B^2 \right)^2}$ \\ \hline
    \end{longtable}
    
\section{Scalar bulk currents}
\label{appendix:scalarcurrent}

{\ Finally there exist radially conserved bulk currents corresponding to the fluctuation of the scalar. These take the form
    \begin{eqnarray}
        \delta \mathcal{I}_{x}(r) &=& \frac{Y[\phi(r)] g(r)}{r^2} f(r) \left( \chi_{x}'(r) - \frac{r k}{g(r)} \tilde{h}_{x}(r) \right) + k n_{f} \bar{E}_{x} I_{Y}(r) \; , \\
        \delta \mathcal{I}_{y}(r) &=& \frac{Y[\phi(r)] g(r)}{r^2} f(r) \left( \chi_{y}'(r) - \frac{r k}{g(r)} \tilde{h}_{y}(r) \right) + k n_{f} \bar{E}_{y} I_{Y}(r) \; ,
    \end{eqnarray}
in the explicit case. In this case these currents tend to something proportional to the vev of the axion fields at the boundary. To obtain the spontaneous case one replaces $\chi_{i}(r) \rightarrow \chi_{i}(r)/r$. In this latter case these currents tend to something proportional to the source of the Goldstone field as $r \rightarrow 0$. In either case the bulk currents associated with the axions/Goldstone bosons are not independent of the other bulk radial currents.}

\section{Numerical solutions}
\label{app:numerics}
In this appendix we detail the numerical simulations leading to the results presented
in section \ref{sec:holomodel} . We will first describe how to construct the background
geometries which are very similar to those at vanishing magnetic field studied
previously in \cite{Amoretti:2018tzw,Amoretti_2019}.
Next we will  study the fluctuations around those backgrounds in order to
characterize the transport properties of the dual system.

\subsection{Black hole geometries}
The holographic model considered in section \ref{sec:holomodel}
admits black hole geometries of the form \eqref{Eq:BackgroundAnsatz1}. In order to construct numerical solutions  within that ansatz we will integrate the equations of motion resulting from \eqref{eq:holoact}  from the horizon at $r=r_h$
to the boundary at $r=0$.
One can easily find the following near-horizon solution for our problem
\begin{subequations}
\begin{align}
&\phi(r)=\phi_h+O(r_h-r)\,,\qquad 
a(r)=a_{h,1}(r_h-r)+O((r_h-r)^2)\,,\\
&f(r)=f_{h,1}(r_h-r)+O((r_h-r)^2)\,,\quad
g(r)=g_h+g_{h,1}(r_h-r)+O((r_h-r)^2)\,,
\end{align}
\label{eq:irsol}
\end{subequations}
where for the potentials \eqref{eq:holopots} one has
\begin{equation}
f_{h,1}={e^{-{\phi_h\over\sqrt{3}}}
	\left[g_h^2\left(6+6 e^{2\phi_h\over\sqrt{3}}
	-r_h^4\,a_{h,1}^2)\right)-B^2r_h^4
	\right]
	-2k^2r_h^2\,g_h\left(1-e^{\phi_h}\right)^2
	\over 2r_h\,g_h\left(2g_h+g_{h,1}\,r_h\right)}\,,
\label{eq:fprimehor}
\end{equation}
which determines the temperature of the black hole $T=-f_{h,1}/(4\pi)$.
All higher order coefficients in \eqref{eq:irsol} are fixed in terms of
$\phi_h$, $a_{h,1}$, $g_h$, and $g_{h,1}$.

One can easily check that the equations of motion are invariant under the scaling
transformation $(t,x,y,r)\to \alpha\,(t,x,y,r)$, $A_t\to A_ t/\alpha$,
$B\to B/\alpha^2$, and 
$k\to k/\alpha$, which we use to set $r_h=1$ in the following.
We can now obtain numerical solutions by integrating the equations
from the horizon ($r=1$) to the boundary ($r=0$).
However, a generic solution will not feature the UV asymptotics \eqref{Eq:UVexpansionbackgroundexp}},
but will behave as $g(r)=g_0+g_1\,r+\ldots$, and one would need to
tune the horizon parameters to look for solutions where $g_0=1$, and $g_1=0$.
Instead we can use  a second scale invariance of the equations under
$(x,y)\to \beta(x,y)$, $k\to k/\beta$, $B\to B/\beta^2$, $g\to g/\beta^2$,
to set $g_0=1$.
This means we can fix $g_h$ and
 we are left with three horizon parameters,
$\phi_h$, $a_{h,1}$, $g_{h,1}$, and one boundary condition $g_0=1$.
Therefore, upon fixing the value of  $B$ and $k$ we expect to obtain a two-parameter family
of solutions which we can parametrize in terms of the two dimensionless
ratios $\lambda/\mu$ and $T/\mu$.

\subsection{Time-dependent fluctuations}
In order to compute the AC correlators studied in section \ref{sec:holomodel} we study the following
 set of fluctuations
 \begin{subequations}
 	\label{eq:fluctuations}
\begin{align}
&\delta g_{tx}=h_x(r)\,e^{-i\omega t}\,,\quad
\delta A_x=a_x(r)\,e^{-i\omega t}\,,\quad
\delta \psi_x=\xi_x(r)\,e^{-i\omega t}\,,\\
&\delta g_{ty}=h_y(r)\,e^{-i\omega t}\,,\quad
\delta A_y=a_y(r)\,e^{-i\omega t}\,,\quad
\delta \psi_y=\xi_y(r)\,e^{-i\omega t}\,.
\end{align}
\end{subequations}
At linear order, the equations of motion resulting from \eqref{eq:holoact} form
a consistent set of six coupled second order differential equations.

Naturally the boundary asymptotics of the fluctuations depend on the background
corresponding to a spontaenous ($\lambda=0$) or explicit ($\lambda\neq0$) solution.
In the spontaneous case they read
 \begin{subequations}
 \label{eq:uvflucssp}
\begin{align}
&h_i(r)=r^{-2}\left[h_{i,0}+h_{i,3}\,r^3+O(r^4)\right]\,,\\
&a_i(r)=a_{i,0}+a_{i,1}\,r+O(r^2)\,,\\
&\xi_i(r)=\xi_{i,-1}/r+\xi_{i,0}+O(r)
\end{align}
\end{subequations}
where $i=x,y$, and $h_{i,3}$ takes the form
\begin{subequations}
\label{eq:uvflucssph3}
\begin{align}
&h_{x,3}={i\over3\omega}
\left[-i\omega\,n_{f}\,a_{x,0}+k\,\xi_{x,-1}\,\phi_v^2+B\left(a_{y,1}-n_{f}\,h_{y,0}
\right)
\right]\,,\\
&h_{y,3}={i\over3\omega}
\left[i\omega\,n_{f}\,a_{y,0}-k\,\xi_{y,-1}\,\phi_v^2+B\left(a_{x,1}-n_{f}\,h_{x,0}
\right)
\right]\,.
\end{align}
\end{subequations}

The boundary behavior of the fluctuations in the explicit case is given by
 \begin{subequations}
	\label{eq:uvflucsxp}
	\begin{align}
	&h_i(r)=r^{-2}\left[h_{i,0}-{1\over4}\lambda\,h_{i,0}\,r^2
	+h_{i,3}\,r^3+O(r^4)\right]\,,\\
	&a_i(r)=a_{i,0}+a_{i,1}\,r+O(r^2)\,,\\
	&\xi_i(r)=\xi_{i,0}+\xi_{i,1}\,r+O(r^2)\,,
	\end{align}
\end{subequations}
with
\begin{subequations}
	\label{eq:uvflucsxph3}
	\begin{align}
	&h_{x,3}={i\over3\omega}
	\left[i\omega\,(-n_{f}\,a_{x,0}+\lambda\,\phi_v\,h_{x,0})-k\,\lambda^2\,\xi_{x,1}
	+B\left(a_{y,1}-n_{f}\,h_{y,0}\right)\right]\,,\\
	&h_{y,3}={i\over3\omega}
	\left[i\omega\,(-n_{f}\,a_{y,0}+\lambda\,\phi_v\,h_{y,0})-k\,\lambda^2\,\xi_{y,1}
	-B\left(a_{x,1}-n_{f}\,h_{x,0}\right)\right]\,.
	\end{align}
\end{subequations}

Since we want to compute the retarded correlators we impose ingoing boundary
conditions for the fluctuations at the horizon:
\begin{subequations}
	\label{eq:irflucs}
	\begin{align}
	&h_i(r)\,(1-r)^{i\omega\over f_{h,1}}=h_{i}^{(h,1)}(1-r)+O((1-r)^2)
	\,,\\
	&a_i(r)\,(1-r)^{i\omega\over f_{h,1}}=a_i^{(h,0)}+O(1-r)\,,\\
	&\xi_i(r)\,(1-r)^{i\omega\over f_{h,1}}=\xi_{i}^{(h,0)}+O(1-r)\,,
	\end{align}
\end{subequations}
with
\begin{subequations}
\begin{align}
&h_x^{(h,1)}={g_h\,k\,\xi_x^{(h,0)}(1-e^{\phi_h})^2
	-e^{-{\phi_h\over\sqrt{3}}}\left(n_{f}\,g_h\,a_x^{(h,0)}+B\,a_y^{(h,0)}\right)
	\over g_h(i\omega/f_{h,1}-1)}\,,\\
&h_y^{(h,1)}={g_h\,k\,\xi_y^{(h,0)}(1-e^{\phi_h})^2
	-e^{-{\phi_h\over\sqrt{3}}}\left(n_{f}\,g_h\,a_y^{(h,0)}-B\,a_x^{(h,0)}\right)
	\over g_h(i\omega/f_{h,1}-1)}\,,
\end{align} 
\end{subequations}
and all other higher order coefficients in \eqref{eq:irflucs} fixed in terms
of $a_i^{(h,0)}$ and $\xi_i^{(h,0)}$ too.

Finally, in order to compute the correlators we will make use of the following
solutions of the equations of motion
\begin{subequations}
	\label{eq:puregaugesols}
\begin{align}
&h_x(r)=i\omega{g(r)\over r^2}\,,\quad h_y(r)=0\,,\quad a_x(r)=0\,,\quad
a_y(r)=B\,,\quad \xi_x=-k\,,\quad \xi_y=0\,,\\
&h_x(r)=0\,,\quad h_y(r)=-i\omega{g(r)\over r^2}\,,\quad a_x(r)=B\,,\quad
a_y(r)=0\,,\quad \xi_x=0\,,\quad \xi_y=k\,,
\end{align}
\end{subequations}
which result from diffeomorphism transformations of the trivial solution.

\subsubsection{AC Correlators}

We shall now compute the retarded correlators $G^R_{AB}$ where 
$A,B=h_x,h_y,a_x,a_y,\xi_x,\xi_y$. We follow~\cite{Kaminski:2009dh} and write
\begin{equation}
G^R_{AB}={\cal B}+{\cal A}\,.\,V\,.\,S^{-1}\,,
\end{equation}
where the matrices ${\cal A}$ and ${\cal B}$ can be extracted from the quadratic action
once it is written as
\begin{equation}
S_{os}^{(2)}=\int d\omega \left(A_{IJ}\,{\mathfrak v}^J\,{\mathfrak s}^I+
B_{IJ}\,{\mathfrak s}^J\,{\mathfrak s}^I\right)\,.
\end{equation}
${\mathfrak s}$ and ${\mathfrak v}$ are vectors made of the leading (sources)
and subleading (vevs) coefficients of the fluctuations at the boundary.
Hence in the spontaneous case we have
\begin{equation}
{\mathfrak s}=(h_{x,0},a_{x,0},\xi_{x,-1},h_{y,0},a_{y,0},\xi_{y,-1})\,,\quad
{\mathfrak v}=(h_{x,3},a_{x,1},\xi_{x,0},h_{y,3},a_{y,1},\xi_{y,0})\,,
\end{equation}
in terms of the coefficients in  \eqref{eq:uvflucssp}, \eqref{eq:uvflucssph3}.
Whereas in the explicit scenario ${\mathfrak s}$ and ${\mathfrak v}$ are
\begin{equation}
{\mathfrak s}=(h_{x,0},a_{x,0},\xi_{x,0},h_{y,0},a_{y,0},\xi_{y,0})\,,\quad
{\mathfrak v}=(h_{x,3},a_{x,1},\xi_{x,1},h_{y,3},a_{y,1},\xi_{y,1})\,,
\end{equation}
now in terms of the coefficients in  \eqref{eq:uvflucsxp}, \eqref{eq:uvflucsxph3}.

By computing the quadratic on-shell action (see e.g. \cite{Amoretti:2020ica} for a derivation) one can check that in the spontaneous case ${\cal A}$ and ${\cal B}$ take the form
\begin{equation}
{\cal A}=\left(
\begin{array}{cccccc}
-3 & 0 & 0 & 0 & 0 & 0\\
0 & 1 & 0 & 0 & 0 & 0 \\
0 & 0 & \phi_v^2 & 0 & 0 & 0\\
0 & 0 & 0 & -3 & 0 & 0\\
0 & 0 & 0 & 0 & 1 & 0 \\
0 & 0 & 0 & 0 & 0 & \phi_v^2\\
\end{array}
\right)\,,\quad
{\cal B}=\left(
\begin{array}{cccccc}
\epsilon_{f}/2 & 0 & 0 & 0 & 0 & 0\\
-n_{f} & 0 & 0 & 0 & 0 & 0 \\
0 & 0 & 0 & 0 & 0 & 0\\
0 & 0 & 0 & \epsilon_{f}/2 & 0 & 0\\
0 & 0 & 0 & -n_{f} & 0 & 0 \\
0 & 0 & 0 & 0 & 0 & 0\\
\end{array}
\right)\,,
\end{equation}
while in the explicit regime ${\cal B}$ is unchanged and ${\cal A}$ reads
\begin{equation}
{\cal A}=\left(
\begin{array}{cccccc}
-3 & 0 & 0 & 0 & 0 & 0\\
0 & 1 & 0 & 0 & 0 & 0 \\
0 & 0 & \lambda^2 & 0 & 0 & 0\\
0 & 0 & 0 & -3 & 0 & 0\\
0 & 0 & 0 & 0 & 1 & 0 \\
0 & 0 & 0 & 0 & 0 & \lambda^2\\
\end{array}
\right)\,.
\end{equation}

Finally, to construct the matrices $S$ and $V$
we need to generate six independent solutions for the fluctuation fields
$h_x, h_y, a_x, a_y, \xi_x, \xi_y$. Two of them are the pure gauge
solutions \eqref{eq:puregaugesols}, and we generate another four solutions
by numerically integrating the equations of motion.
As is clear from \eqref{eq:irflucs}, there are four free parameters at the
horizon that allow us to generate four independent solutions.
$S$ and $V$ are respectively the matrices of sources and vevs constructed
out of six independent solutions as
$S=({\mathfrak s}^{I},{\mathfrak s}^{II},{\mathfrak s}^{III},
{\mathfrak s}^{VI},{\mathfrak s}^{V},{\mathfrak s}^{VI})$,
$V=({\mathfrak v}^{I},{\mathfrak v}^{II},{\mathfrak v}^{III}
{\mathfrak v}^{IV},{\mathfrak v}^{V},{\mathfrak v}^{VI})$. They read
\begin{equation}
\label{eq:SVmatricessp}
S=\left(
\begin{array}{cccccc}
h_{x,0}^{(I)} & h_{x,0}^{(II)} & h_{x,0}^{(III)} & h_{x,0}^{(IV)} & 0
&i\omega \\
a_{x,0}^{(I)} & a_{x,0}^{(II)} & a_{x,0}^{(III)} & a_{x,0}^{(IV)} & B
&0 \\
\xi^{(I)}_{x,-1} & \xi^{(II)}_{x,-1} & \xi^{(III)}_{x,-1}  & \xi^{(IV)}_{x,-1} 
&0 &0\\
h_{y,0}^{(I)} & h_{y,0}^{(II)} & h_{y,0}^{(III)} & h_{y,0}^{(IV)} & -i\omega
&0 \\
a_{y,0}^{(I)} & a_{y,0}^{(II)} & a_{y,0}^{(III)} & a_{y,0}^{(IV)} & 0
&B \\
\xi^{(I)}_{y,-1} & \xi^{(II)}_{y,-1} & \xi^{(III)}_{y,-1}  & \xi^{(IV)}_{y,-1} 
&0 &0
\end{array}
\right)\,,\quad
V=\left(
\begin{array}{cccccc}
h_{x,3}^{(I)} & h_{x,3}^{(II)} & h_{x,3}^{(III)} & h_{x,3}^{(IV)} & 0
&0 \\
a_{x,1}^{(I)} & a_{x,1}^{(II)} & a_{x,1}^{(III)} & a_{x,1}^{(IV)} & 0
&0 \\
\xi^{(I)}_{x,0} & \xi^{(II)}_{x,0} & \xi^{(III)}_{x,0}  & \xi^{(IV)}_{x,0} 
&0 &-k\\
h_{y,3}^{(I)} & h_{y,3}^{(II)} & h_{y,3}^{(III)} & h_{y,3}^{(IV)} &0
&0 \\
a_{y,1}^{(I)} & a_{y,1}^{(II)} & a_{y,1}^{(III)} & a_{y,1}^{(IV)} & 0
&0 \\
\xi^{(I)}_{y,0} & \xi^{(II)}_{y,0} & \xi^{(III)}_{y,0}  & \xi^{(IV)}_{y,0} 
&k &0
\end{array}
\right)\,,
\end{equation}
for the spontaneous case. For the explicit regime we have
\begin{equation}
\label{eq:SVmatricesxp}
S=\left(
\begin{array}{cccccc}
h_{x,0}^{(I)} & h_{x,0}^{(II)} & h_{x,0}^{(III)} & h_{x,0}^{(IV)} & 0
&i\omega \\
a_{x,0}^{(I)} & a_{x,0}^{(II)} & a_{x,0}^{(III)} & a_{x,0}^{(IV)} & B
&0 \\
\xi^{(I)}_{x,0} & \xi^{(II)}_{x,0} & \xi^{(III)}_{x,0}  & \xi^{(IV)}_{x,0} 
&0 &-k\\
h_{y,0}^{(I)} & h_{y,0}^{(II)} & h_{y,0}^{(III)} & h_{y,0}^{(IV)} & -i\omega
&0 \\
a_{y,0}^{(I)} & a_{y,0}^{(II)} & a_{y,0}^{(III)} & a_{y,0}^{(IV)} & 0
&B \\
\xi^{(I)}_{y,0} & \xi^{(II)}_{y,0} & \xi^{(III)}_{y,0}  & \xi^{(IV)}_{y,0} 
&k &0
\end{array}
\right)\,,\quad
V=\left(
\begin{array}{cccccc}
h_{x,3}^{(I)} & h_{x,3}^{(II)} & h_{x,3}^{(III)} & h_{x,3}^{(IV)} & 0
&0 \\
a_{x,1}^{(I)} & a_{x,1}^{(II)} & a_{x,1}^{(III)} & a_{x,1}^{(IV)} & 0
&0 \\
\xi^{(I)}_{x,1} & \xi^{(II)}_{x,1} & \xi^{(III)}_{x,1}  & \xi^{(IV)}_{x,1} 
&0 &0\\
h_{y,3}^{(I)} & h_{y,3}^{(II)} & h_{y,3}^{(III)} & h_{y,3}^{(IV)} &0
&0 \\
a_{y,1}^{(I)} & a_{y,1}^{(II)} & a_{y,1}^{(III)} & a_{y,1}^{(IV)} & 0
&0 \\
\xi^{(I)}_{y,1} & \xi^{(II)}_{y,1} & \xi^{(III)}_{y,1}  & \xi^{(IV)}_{y,1} 
&0 &0
\end{array}
\right)\,.
\end{equation}

\subsubsection{Quasinormal Modes}

We are interested in computing the QNMs at zero momentum in the sector given by the fluctuations \eqref{eq:fluctuations}. These are the frequencies at which the 
retarded correlators $G^R_{AB}(\omega)$ (with $A,B=h_x,h_y,a_x,a_y,\xi_x,\xi_y$) have
poles. Again we folow~\cite{Kaminski:2009dh}  and employ the so-called determinant
method. The QNMs are given by the values of $\omega$ for which the
determinant of the matrix of sources $S$ in \eqref{eq:SVmatricessp} or \eqref{eq:SVmatricesxp}  vanishes.

\bibliography{references}
\bibliographystyle{jhep}

\end{document}